\documentclass[12pt,preprint]{aastex}

\def\kms{$\,$km$\,$s$^{-1}$}
\def\sqig{$\sim$}
\def\sun{$_\odot$}
\def\cts{counts~s$^{-1}$}

\def\ergss{ergs s$^{-1}$}

\def\src{4U\,1954+31}
\def\19{4U\,1954+31}
\def\17{4U\,1700+24}
\def\gx{GX\,1+4}

\begin{document}

\title{A Comparison of the Variability of the Symbiotic X-ray Binaries
GX 1+4, 4U 1954+31, and 4U 1700+24 from Swift/BAT and RXTE/ASM
Observations}


\author{R.~H.~D.~Corbet\altaffilmark{1, 2},
J.~L.~Sokoloski\altaffilmark{3},
K.~Mukai\altaffilmark{1, 2},
C.~B.~Markwardt\altaffilmark{1,4},
and J.~Tueller\altaffilmark{5}}

\altaffiltext{1}
{X-ray Astrophysics Laboratory, Mail Code 662,
NASA Goddard Space Flight Center, Greenbelt, MD 20771;
Robin.Corbet@nasa.gov}

\altaffiltext{2}{CRESST; University of Maryland, Baltimore
County}

\altaffiltext{3}{Columbia Astrophysics Laboratory, 550 W120th St.,
1027 Pupin Hall,
Columbia University, NY 10027.}

\altaffiltext{4}{CRESST; Department of Astronomy, University of Maryland,
College Park}

\altaffiltext{5}{Astroparticle Physics Laboratory,
Mail Code 661, NASA/Goddard Space Flight Center, Greenbelt, MD 20771.}

\begin{abstract}
We present an analysis of the X-ray variability of three
symbiotic X-ray binaries, GX 1+4, \17, and \src, using observations
made with the Swift Burst Alert Telescope (BAT) and the Rossi X-ray
Timing Explorer (RXTE) All-Sky Monitor (ASM).  Observations of \src\
with the Swift BAT show modulation at a period near 5 hours. Models to
explain this modulation are discussed including the presence of an
exceptionally slow X-ray pulsar in the system and accretion
instabilities.  We conclude that the most likely
interpretation is that \src\ contains one of the slowest
known X-ray pulsars.
Unlike \src, neither GX 1+4 nor \17\ show any evidence
for modulation on a timescale of hours.  An analysis of the 
RXTE ASM light curves of GX 1+4, \17, and \src\ does not
show the presence of periodic modulation in any source, although there
is considerable variability on long timescales for all three
sources.  There is no
modulation in GX 1+4 on either the optical 1161 day orbital period
or a previously reported 304 day X-ray period. For
\17\ we do not confirm the 404 day X-ray period previously proposed for this
source from a shorter duration ASM light curve.
We conclude that all three sources have substantial low-frequency
noise in their power spectra that may give the appearance of
periodic modulation if this noise is not properly accounted for,
particularly if short duration light curves are examined.

\end{abstract}

\section{Introduction}

A symbiotic binary is a system which contains a
hot object accreting from an M
giant companion,
either from the wind of the M star or via Roche lobe overflow 
(e.g. Iben \& Tutukov 1996).
In most symbiotic stars the accreting object
is a white dwarf, and these systems can be modest X-ray emitters (e.g.
M\"urset et al. 1997). 
In far fewer systems is the accreting object
thought to be a neutron star. There are presently only five
sources for which there is strong evidence that
they are ``symbiotic X-ray binaries'' with
a neutron star component. This paper discusses
three of these sources:  GX 1+4, \src, and \17.
GX 1+4 has been known as a symbiotic
source containing an X-ray pulsar
for some time (e.g. Davidsen et al. 1977). 
\17\ was suspected to be a symbiotic source and this is
now confirmed and \src, 
which was initially thought to be a high-mass
X-ray binary, has also recently been found to have
an M giant counterpart (Masetti et al. 2006). 
In addition to these sources, Sct X-1
IGR J16194-2810, and 1RXS J180431.1-273932 have also recently been suggested to
be symbiotic X-ray binaries (Kaplan et al. 2007, Masetti et al.
2007b, Nucita et al. 2007).

We present an analysis of the
X-ray variability of GX 1+4, \src, and \17
using long term
observations made with the Swift Burst Alert Telescope (BAT)
and the Rossi X-ray Timing Explorer (RXTE) All Sky Monitor (ASM).
Sct X-1, IGR J16194-2810, and
1RXS J180431.1-273932 are not considered here as these sources
were not strongly detected by the BAT
and ASM light curves are not available for IGR J16194-2810
and 1RXS J180431.1-273932.
These observations enable a study of variability on timescales
from hours to years.
The BAT observations
show the presence of a strong modulation with period of
about 5 hours in \src, first reported by Corbet et al. (2006).
We consider possible mechanisms that could cause this modulation.
Modulation on timescales of hours is not seen in either \gx\ or
\17.
We find no evidence for orbital modulation in any of these
three systems.

\section{Previous Observations of Symbiotic X-ray Binaries}

We summarize here previous observations of 
symbiotic X-ray binaries with emphasis on variability and previously claimed
periodicities in \gx, \src, and
\17. Source parameters are listed in Table 1.

\subsection{GX 1+4}

Pulsations with a period
of \sqig120s from GX 1+4 were 
discovered by Lewin et al. (1971).
The optical counterpart was identified as the bright
infrared source V2116 Oph (Glass \& Feast 1973)
and GX 1+4
was classified
as a symbiotic X-ray binary 
containing an M giant mass donor
by Davidsen et al. (1977), making it the first such object to be identified.
The long term X-ray light curve shows large variability,
and the pulsar shows both spin-up and spin-down (e.g. Chakrabarty
et al. 1997).
Chakrabarty \& Roche (1997) presented extensive optical,
infrared, and X-ray observations of GX 1+4 and concluded
that the binary period must be greater than 100 days, and is
most likely greater than 260 days, based on the assumption that
the M giant does not overfill its Roche lobe.
Cutler et al. (1986) proposed
an orbital period of approximately 304 days
based on variations in the pulse period measured
with the high-energy X-ray spectrometer on OSO-8,
and Pereira et al. (1999, 2000)
claimed confirmation for this from BATSE pulse
period measurements
with a refined period of 303.8 $\pm$ 1.1 days.
Pereira et al. (1999) noted that the pulse period variations
could not be attributed to orbital Doppler modulation
as this would require an implausibly large mass for
the companion star and instead the period changes must be related
to changes in the accretion torque.  
Using infrared radial velocity measurements
of the M giant, Hinkle et al. (2006) excluded
the 304 day period as an orbital period and instead found a
period of 1161 $\pm$ 12 days and a system
eccentricity of 0.10 $\pm$ 0.02. 
GX 1+4 is thus a very variable source with well
studied pulsations. Its classification as
a symbiotic X-ray binary is firm -
the pulsations, the change in their period,
and the high luminosity clearly show that the
system contains a neutron star.

\subsection{\src\ (3A 1954+319)}

Pointed observations of \src\  were first
obtained with
EXOSAT (Cook et al. 1984) and
Ginga (Tweedy et al. 1989).
In both sets of observations flaring behavior on
timescales of minutes was reported.
Cook et al. (1984) found that the X-ray spectrum could be fitted
with either a power-law or a thermal bremsstrahlung
model. Tweedy et al. (1989) reported that a more complicated
model was required such as the typical X-ray pulsar
spectrum of a power-law with high energy cutoff. However,
even with this model Tweedy et al. (1989)
found evidence for the presence of an additional soft component.
Because of the variability and the power law spectrum both
Cook et al. (1984) and Tweedy et al. (1989) concluded
that \src\ was likely to be a high-mass X-ray binary (HMXB).
However, Masetti et al. (2006) obtained a precise
position from Chandra observations which enabled the optical
counterpart to be identified as
an M giant. \src\ is thus a symbiotic system rather than
an HMXB. Masetti et al. (2007a) give a summary of a variety
of X-ray observations of \src. 
No orbital period has been proposed for this system
and, apart from the 5 hour period discussed in this paper,
no pulsations had previously been reported.

\subsection{\17\ (2A 1704+241)}
\17\ was proposed to be a symbiotic X-ray binary by
Garcia et al. (1983) on the basis of a positional
association with the M giant star HD 154791 obtained from
observations with the Einstein Observatory Imaging
Proportional Counter (IPC). A position obtained
with the ROSAT  High Resolution Imager by Morgan \& Garcia (2001)
was apparently inconsistent with
HD 154791. However, subsequent observations with the Chandra
High Resolution Camera Imager 
by Masetti et al. (2006) gave a position consistent
with that of HD 154791 and inconsistent with that reported
by Morgan \& Garcia (2001). The Masetti et al. (2006) position
appears to firmly identify \17\ as a symbiotic system.

Garcia et al. (1983) reported the presence of
900s quasi-periodic modulation from their IPC observations.
Morgan \& Garcia (2001) also claimed to detect
modulation near a period of 900s in ROSAT observations,
but with a different frequency from that reported
previously. Masetti et al. (2002)
reported on ASCA, BeppoSAX, ROSAT, and RXTE Proportional
Counter Array (PCA) observations of \17. They found
erratic ``shot noise'' variability on timescales
of tens to thousands of seconds but did not confirm
the modulation on the 900s timescale reported by
Garcia et al. (1983) and Morgan \& Garcia (2001).

Masetti et al. (2002) reported a tentative periodicity
of \sqig400 days from RXTE ASM observations obtained
between 1996 and 2001. Galloway et al. 
(2002) also analyzed RXTE ASM observations
obtained between January 1996 and April 2002
and claimed to confirm this modulation at a
refined period of 404 $\pm$ 20
days. Galloway et al. (2002) also reported optical radial velocity
measurements which showed marginally significant
modulation at a period of 404 $\pm$ 3 days 
and suggested that the modulation could
be caused by either orbital motion or pulsational modulation
of the red giant. 

In summary, there is strong evidence that \17\ is a symbiotic
X-ray source. It is highly variable in the X-ray band,
there is
possible evidence of
modulation on a \sqig400 day timescale, 
but there is no convincing evidence of pulsations.

\subsection{Scutum X-1, IGR J16194-2810, and 1RXS J180431.1-273932}
Sct X-1 has been observed by many instruments since
its discovery in a rocket flight by Hill et al. (1974).
Kaplan et al. (2007) recently derived a corrected position
for Sct X-1 that showed it to be associated with
a late type giant or supergiant optical counterpart.
Pulsations at a period of 111 s were first detected in Ginga observations 
(Makino 1988; Koyama et al. 1991) and
Kaplan et al. (2007) showed 
from observations with Ginga, RXTE, and XMM-Newton
that the pulsar has
exhibited spin-down over a period of 17 years.

IGR J16194-2810 is another recently discovered
symbiotic X-ray binary.
The source was reported in the 2nd IBIS survey (Bird
et al. 2006) and associated with the ROSAT source
1RXS J161933.6-280736 by Stephen et al. (2006).
Masetti et al. (2007b) used
a position obtained with the Swift X-ray Telescope to
identify an M2 III optical counterpart. Masetti et al.
reported erratic X-ray variability on timescales
of hundreds to thousands of seconds, but no pulsations
were found.

1RXS J180431.1-273932 has been found to be a pulsar
with a 494s period by Nucita et al. (2007) from XMM
observations. Nucita et al. (2007) also identify an optical
counterpart with a possible M6III classification. This classification
would make 1RXS J180431.1-273932 another pulsing symbiotic
X-ray binary.

\section{Observations}

\subsection{Swift BAT}
The Swift BAT is described in detail by Barthelmy et al. (2005) 
and the data
reduction process is described by Markwardt et al. (2005)
and Tueller et al. (2007).
The BAT is a very wide field of view (1.4 sr half-coded)
hard X-ray telescope that utilizes
a 2.7 m$^2$ coded-aperture mask and a 0.52 m$^2$ CdZnTe detector array
divided into 32,768 detectors each with an area of 0.16 cm$^2$.
The pointing direction of the BAT is determined by observations
using the narrow-field
XRT and UVOT instruments also on board Swift which are primarily used
to study gamma-ray bursts and their afterglows. BAT observations
of X-ray sources are thus generally obtained in a serendipitous
and unpredictable fashion. Typically the BAT observes 50\% to 80\%
of the sky each day.
The data considered here consists of individual ``snapshots'',
i.e. times spent observing the same position without a break. 
These snapshots
have exposure times ranging between 150 to 2678 s with means
of 891 s (\gx), 987 s (\src) and 979 s (\17).
The time resolution of our data is equal to the snapshot duration.

From these snapshots light curves are constructed in
4 energy bands: 14 to 25, 25 to 50, 
50 to 100, and 100 to 195  keV.
For the entire energy range the Crab produces approximately
0.045 \cts\ per fully illuminated  detector for an equivalent
on-axis source (hereafter abbreviated to \cts).
In each energy band the Crab produces 0.019, 0.018, 0.0077,
and 0.0012 \cts\ respectively.
The BAT light curves used here cover the periods
MJD 53,373 to 53,818 (GX 1+4), 53,350 to 53,815 (\src),
and 53,352 to 53,818 (\17).

\subsection{RXTE ASM}

The RXTE ASM (Levine et al. 1996) consists of three similar
Scanning Shadow Cameras
which perform sets of 90 second pointed
observations (``dwells'') so as to cover \sqig80\% of the sky every
\sqig90 minutes.  
Light curves are available in three energy bands: 1.5 to 3 keV, 3
to 5 keV, and 5 to 12 keV.
The Crab produces approximately 75.5
\cts\ in the ASM over the entire energy range and 
26.8 (1.5 to 3.0 keV), 23.3 (3 to 5 keV), and 25.4 (5 to 12 keV) 
\cts\ in each energy band. 
Observations
of blank field regions away from the Galactic center suggest that
background subtraction may produce a systematic uncertainty of about 0.1
\cts\ (Remillard \& Levine 1997).
The ASM light curves used in our analysis 
span  the period from MJD 50,087 to 54,195 
except for the light curve of \gx\  which starts on MJD 50,088.

\section{Analysis and Results}

The RXTE ASM light curves cover a longer duration than
the Swift BAT light curves and so are better suited
for searches for orbital modulation on the long time scales
expected for symbiotic X-ray binaries.
The BAT light curves are generally more sensitive than
the ASM light curves, particularly for highly-absorbed sources,
and so have advantages in studying variability on shorter
timescales. We use power spectra of the light curves
to search for periodic modulation and
in all cases the calculation of power spectra employed the
``semi-weighting'' scheme discussed in Corbet et al. 
(2007a, b).

\subsection{\gx}

\begin{figure}
\epsscale{0.9}
\plotone{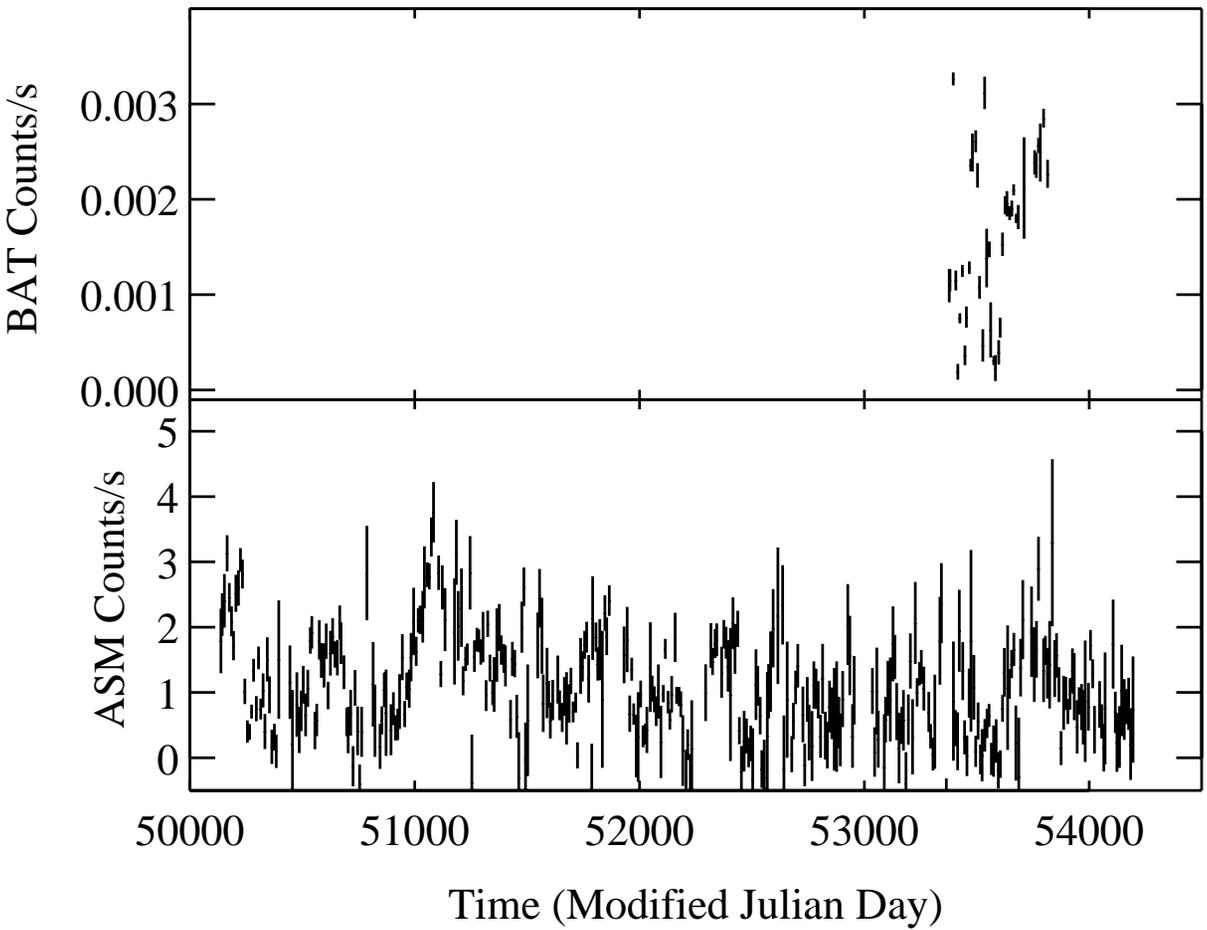}
\figcaption[f1.eps]{BAT and ASM light curves of GX 1+4 in ten day
averages. The full energy ranges are used for each instrument.
\label{fig:gx_lc}
}
\end{figure}

The ASM and BAT light curves of \gx\ are shown in
Fig. \ref{fig:gx_lc} and the mean count rates
are given in Table 1. There is considerable variability seen
with both instruments. During the interval covered simultaneously by
both instruments the BAT light curve shows dramatic
flaring while the modulation in the ASM light curve
appears to have lower amplitude.

\begin{figure}
\epsscale{0.9}
\plotone{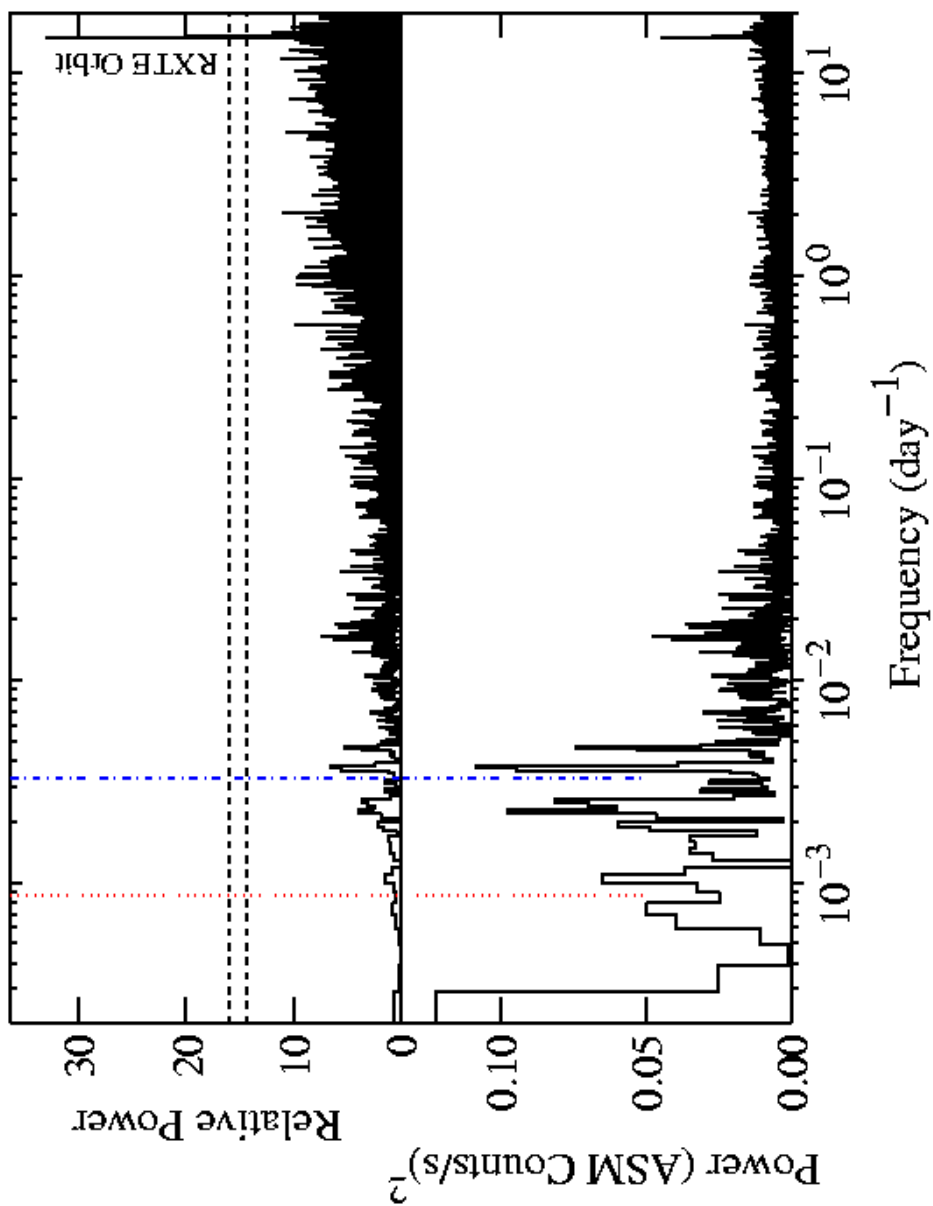}
\figcaption[f2.eps]{Power spectrum of the ASM
light curve of GX 1+4. 
The lower panel shows the uncorrected power spectrum
and the upper panel shows the power spectrum with the
estimated continuum noise component removed.
The horizontal dashed lines in the upper panel indicate
the estimated 95\% and 99\% significance levels.
The 1161 day orbital period
found by Hinkle et al. (2006) is marked by
the vertical dotted red line (left) and the 304 day period
of Cutler et al. (1986) is marked by the vertical dot-dashed blue line (right).
\label{fig:gx_asm_ft}
}
\end{figure}

The power spectrum of the
entire ASM light curve for periods
between 0.05 to 4100 days with 200,000 frequency
steps is shown in Fig. \ref{fig:gx_asm_ft}. 
The length of the light curve implies a frequency resolution
of 2.4$\times$10$^{-4}$ day$^{-1}$ which was confirmed
by calculating power spectra with the data values replaced
with sine waves. 
The power spectrum in Fig. \ref{fig:gx_asm_ft}
is thus oversampled by a factor of 2.4.
Although
there are a number of peaks in the power spectrum at
low frequencies, none exactly coincides with either the 1160 day
period proposed by Hinkle et al. (2006)
or the 304 day period reported by Cutler et al. (1986).
Although the largest peak in the power spectrum is at 270 days,
which is similar length
to the Cutler et al. 304 day period, the peak does not
overlap
with this period.
There are also other peaks almost as strong as
the peak at 270 days, which appears to be part of
the general low frequency noise present in the power spectrum.

\begin{figure}
\epsscale{0.9}
\plotone{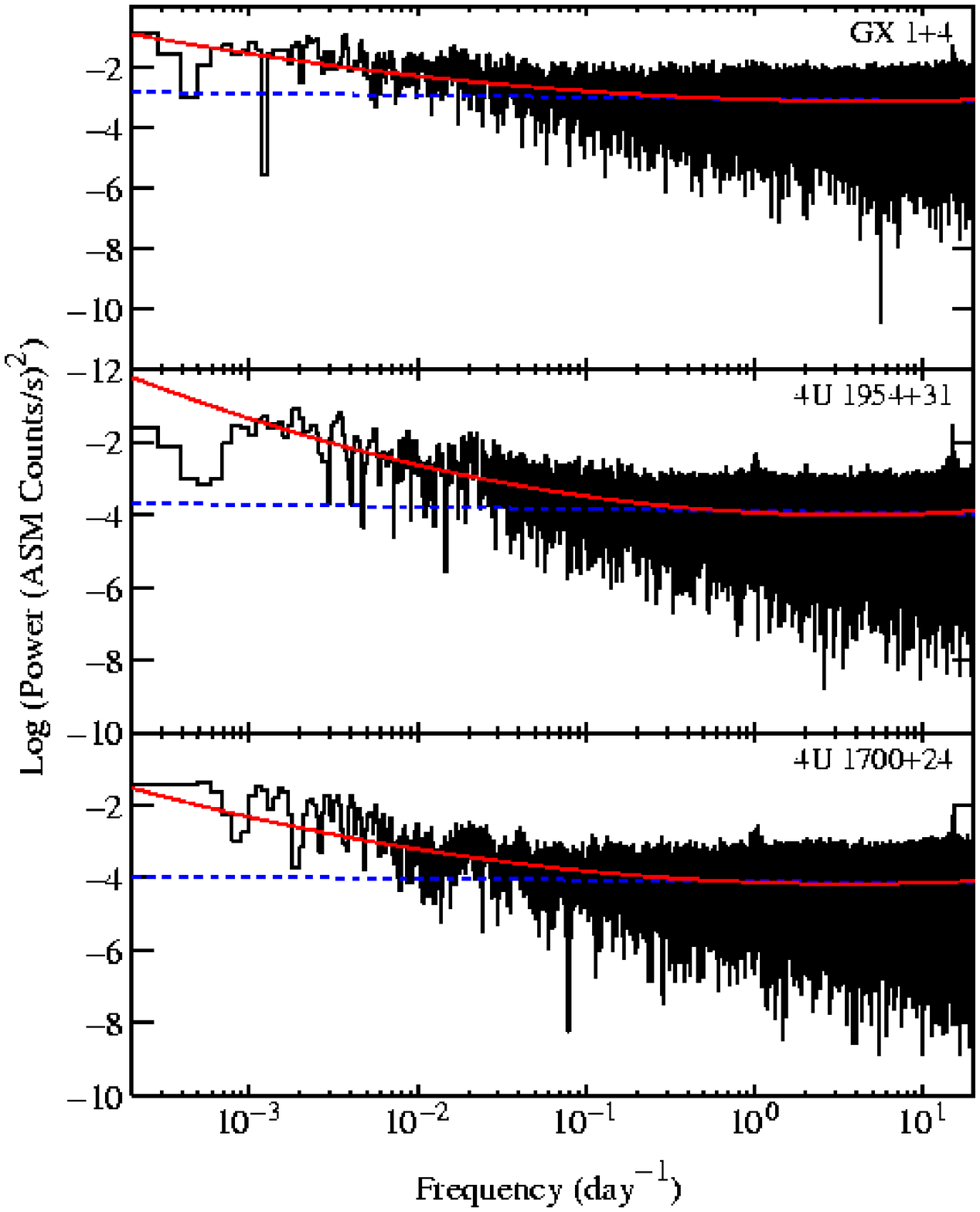}
\figcaption[f3.eps]{Fits to the continuum power spectra
of the ASM light curves of \gx\ (top panel), \src\ (middle panel)
and \17\ (bottom panel). The dashed blue lines show
linear fits to the logarithms of the power spectra
(i.e. power laws) and the solid red lines show quadratic
fits. The quadratic fits were used to remove the low frequency
noise.
\label{fig:noise_fits}
}
\end{figure}

In order to estimate the amount of low frequency noise
present in the spectrum and remove it we examined the
relationship between the continuum power level and frequency.
This was done using the logarithms of power and frequency
as advocated, for example, by Vaughan (2005). It
was found that a linear fit to log(power) vs. log(frequency),
i.e. a power-law relationship between power and frequency,
underestimated the power at low frequencies (Fig.  
\ref{fig:noise_fits}). Instead it was found that a quadratic
fit gave a reasonable approximation to the continuum
noise level. This quadratic fit was subtracted from the
logarithm of the power spectrum together with a constant
of value 0.25068 in order to account for the bias due
to the $\chi^2$ distribution of the power spectrum (e.g.
Papadakis \& Lawrence 1993, Vaughan 2005).
The power spectrum corrected in this way for the ``red noise''
contribution is shown in the upper panel of Fig. \ref{fig:gx_asm_ft}.
The 95\% and 99\% significance levels are also plotted, although
these do not include the effects of the red noise subtraction,
including the choice of model for this. These effects
are greater at lower frequencies and thus even larger
power spectrum levels than those marked would be required
to achieve true 95\% or 99\% significance levels.
With the continuum noise removed in this way it can
be seen that none of the low frequency peaks achieves
statistical significance.
At higher frequencies there is a significant
peak around the orbital period of RXTE at \sqig15.0185 day$^{-1}$
and a smaller peak close to 1 day$^{-1}$ which is likely
related to aliasing of the strong low frequency noise
with a one day sampling pattern (see Farrell, O'Neill
\& Sood 2005, 
Wen et al. 2006).

\begin{figure}
\epsscale{0.9}
\plotone{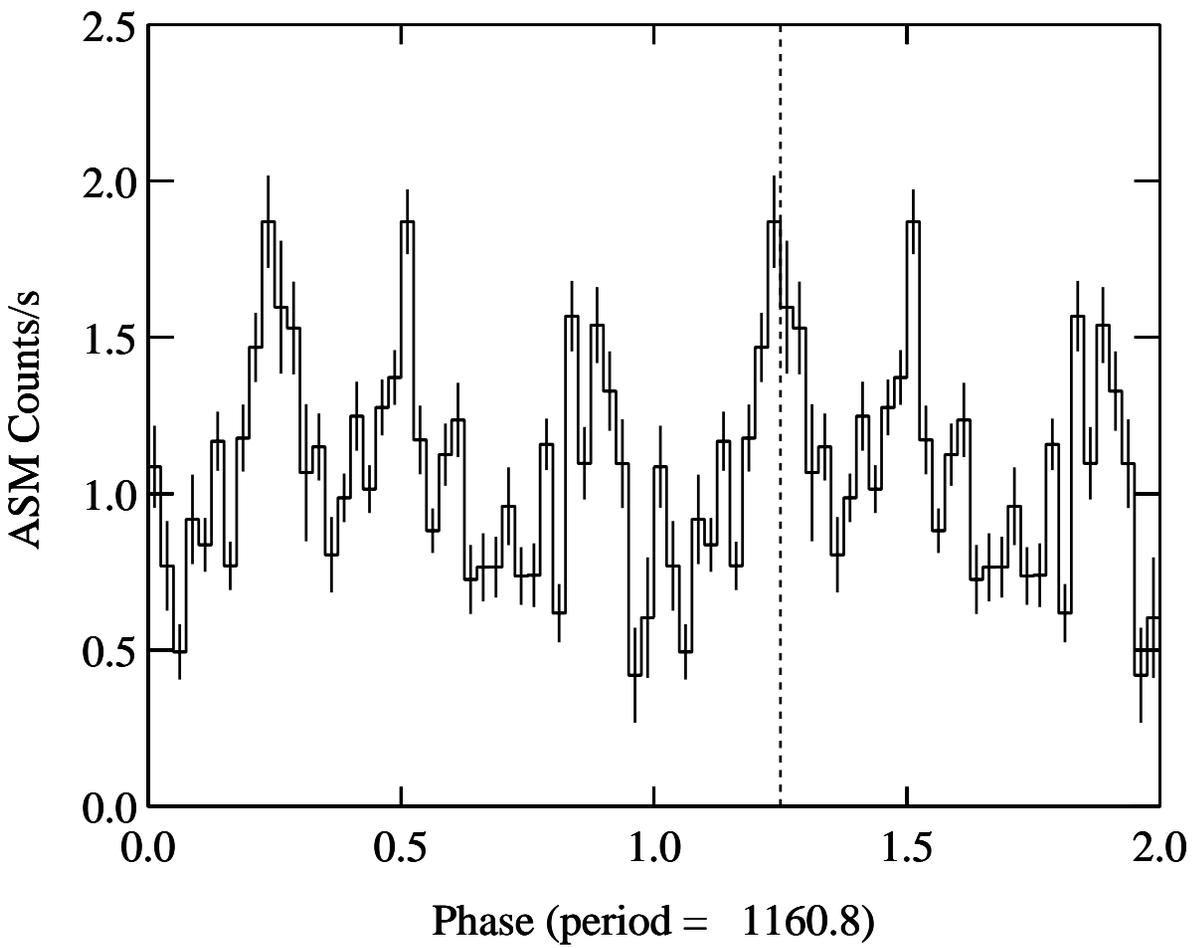}
\figcaption[f4.eps]{The ASM light curve of \gx\ folded
on the 1160 day period from Hinkle et al. (2006).
Phase 0 corresponds to periastron passage and the
dashed line indicates the phase when an eclipse
of the neutron star could take place.
\label{fig:gx_fold}
}
\end{figure}

In Fig. \ref{fig:gx_fold}
we show the ASM light curve of GX 1+4 folded
on the Hinkle et al. (2006) orbital period.
The folded light curve does not show obvious periodic
modulation. In particular, there is neither a flux enhancement
at periastron passage nor flux reduction at the phase of
predicted eclipse (Hinkle et al. 2006).

\begin{figure}
\epsscale{0.9}
\plotone{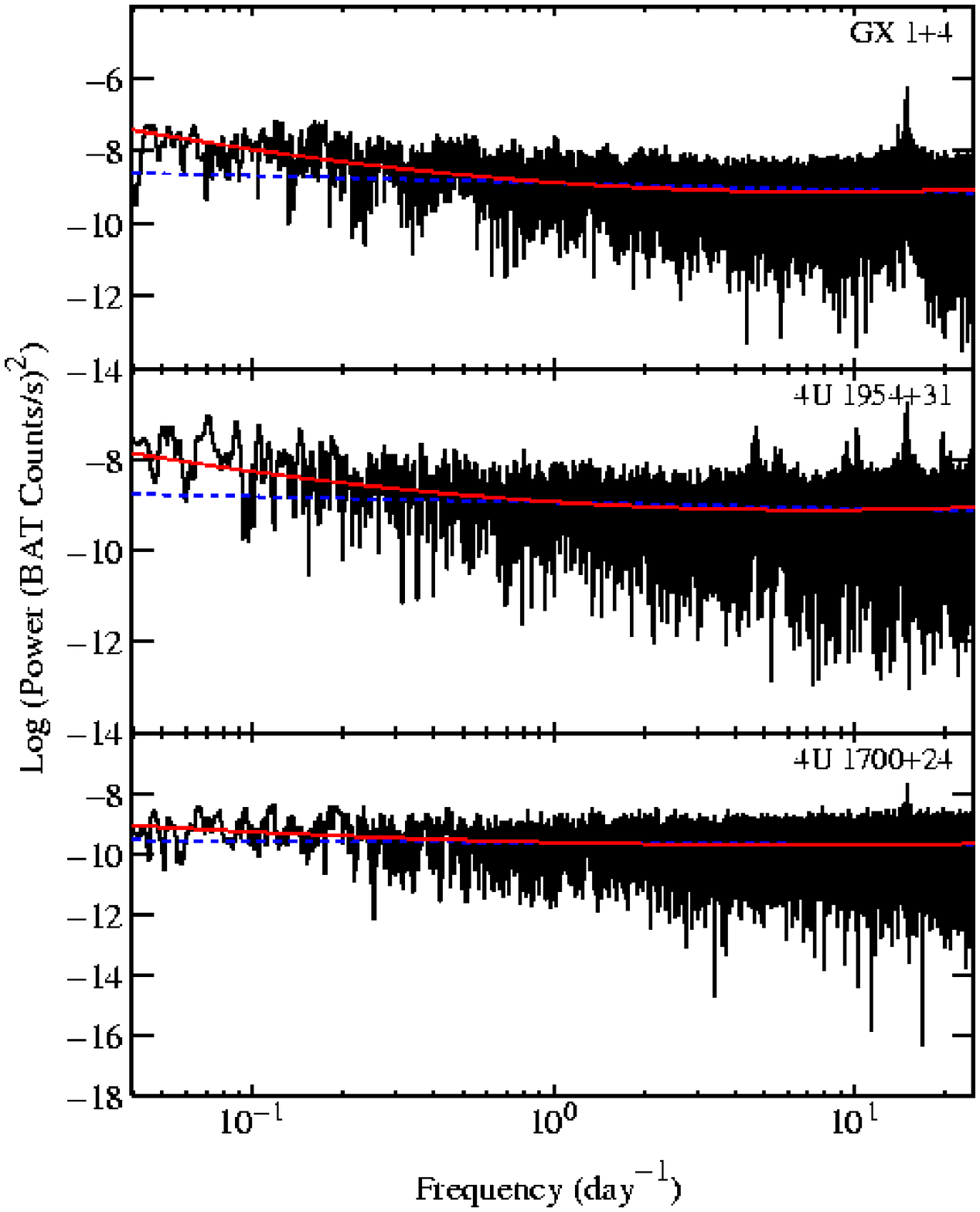}
\figcaption[f5.eps]{Fits to the continuum power spectra
of the BAT light curves of \gx\ (top panel), \src\ (middle panel)
and \17\ (bottom panel). The dashed blue lines show
linear fits to the logarithms of the power spectra
(i.e. power laws) and the solid red lines show quadratic
fits. The quadratic fits were used to remove the low frequency
noise.
\label{fig:bat_noise_fits}
}
\end{figure}

\begin{figure}
\epsscale{0.9}
\plotone{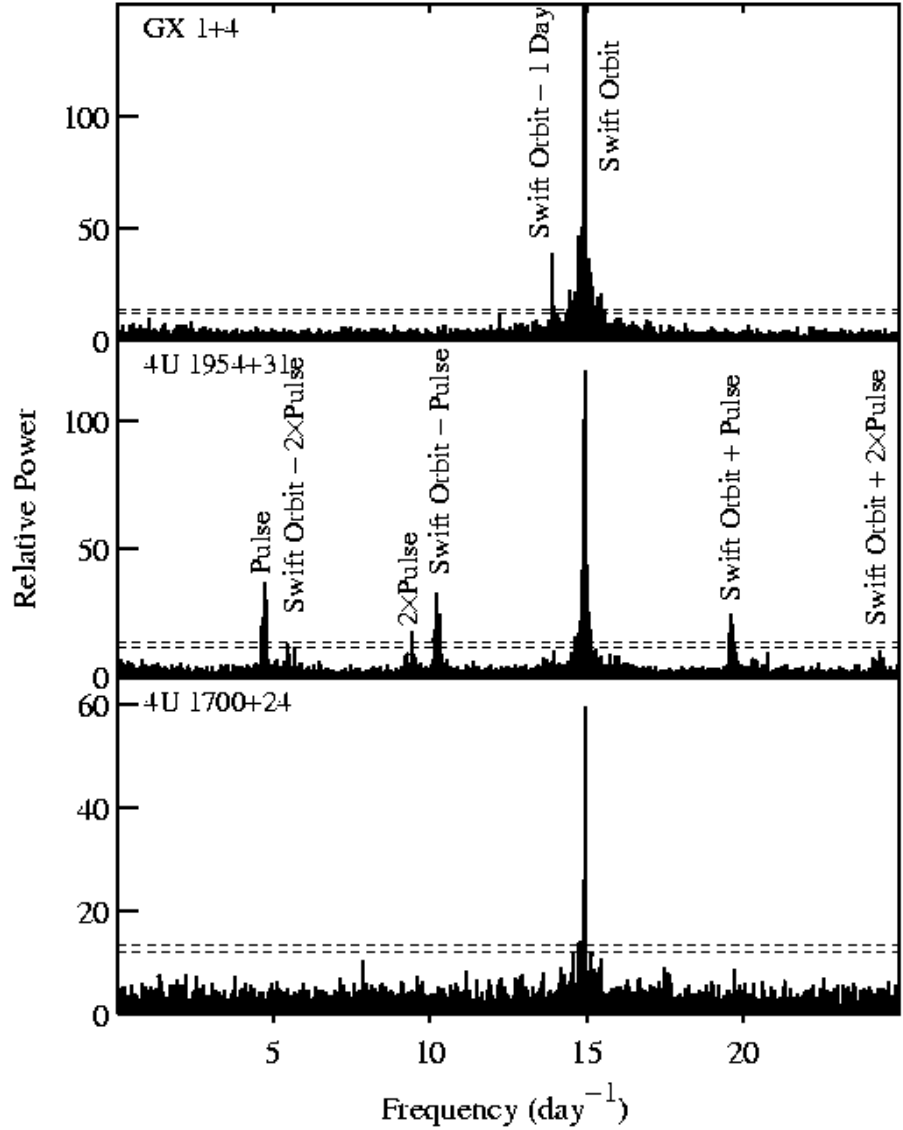}
\figcaption[f6.eps]{Power spectra of BAT light curves.
Upper panel: GX 1+4, middle panel: \src, lower panel: \17.
The power range in the lower panel is truncated to
enable lower amplitude peaks to be seen.
In all cases low frequency noise has been subtracted
using the fits shown in Fig. \ref{fig:bat_noise_fits}.
The horizontal dashed lines in all panels indicate
the estimated 95\% and 99\% significance levels.
\label{fig:bat_ft}
}
\end{figure}

We next investigated the BAT light curve of
\gx\ to search for modulation at higher frequencies.
The BAT power spectrum of \gx\ was calculated for the period
range of 0.04 to 25 days with 40,000 frequency steps.
The total light curve duration
of about 440 days implies a nominal frequency resolution of
about 2$\times$10$^{-3}$ day$^{-1}$. However, there are several
gaps in the light curve and an analysis with
data values replaced with a sine wave indicates that
the resolution is somewhat worse at \sqig2.5$\times$10$^{-3}$ day$^{-1}$.
The calculated power spectrum thus is oversampled by a factor
of \sqig4.
In a similar way that low frequency noise was removed from
the ASM power spectrum we again fitted log(power) vs. log(frequency)
with a quadratic function as shown in Fig. \ref{fig:bat_noise_fits}.
The BAT power spectrum of \gx\ corrected for
the continuum noise level is shown  Fig. \ref{fig:bat_ft}.
No features due to source variability are detected.
A set of peaks in the range of approximately 14.89
to 14.96 day$^{-1}$ 
is
due to the orbital period of
the Swift satellite. At slightly lower frequencies,
another set of small peaks is seen near 13.92  day$^{-1}$.
As these peaks are 1 day$^{-1}$ less than the Swift
orbit peaks, these are also thought to be artifacts.

\subsection{\src}

\begin{figure}
\epsscale{0.9}
\plotone{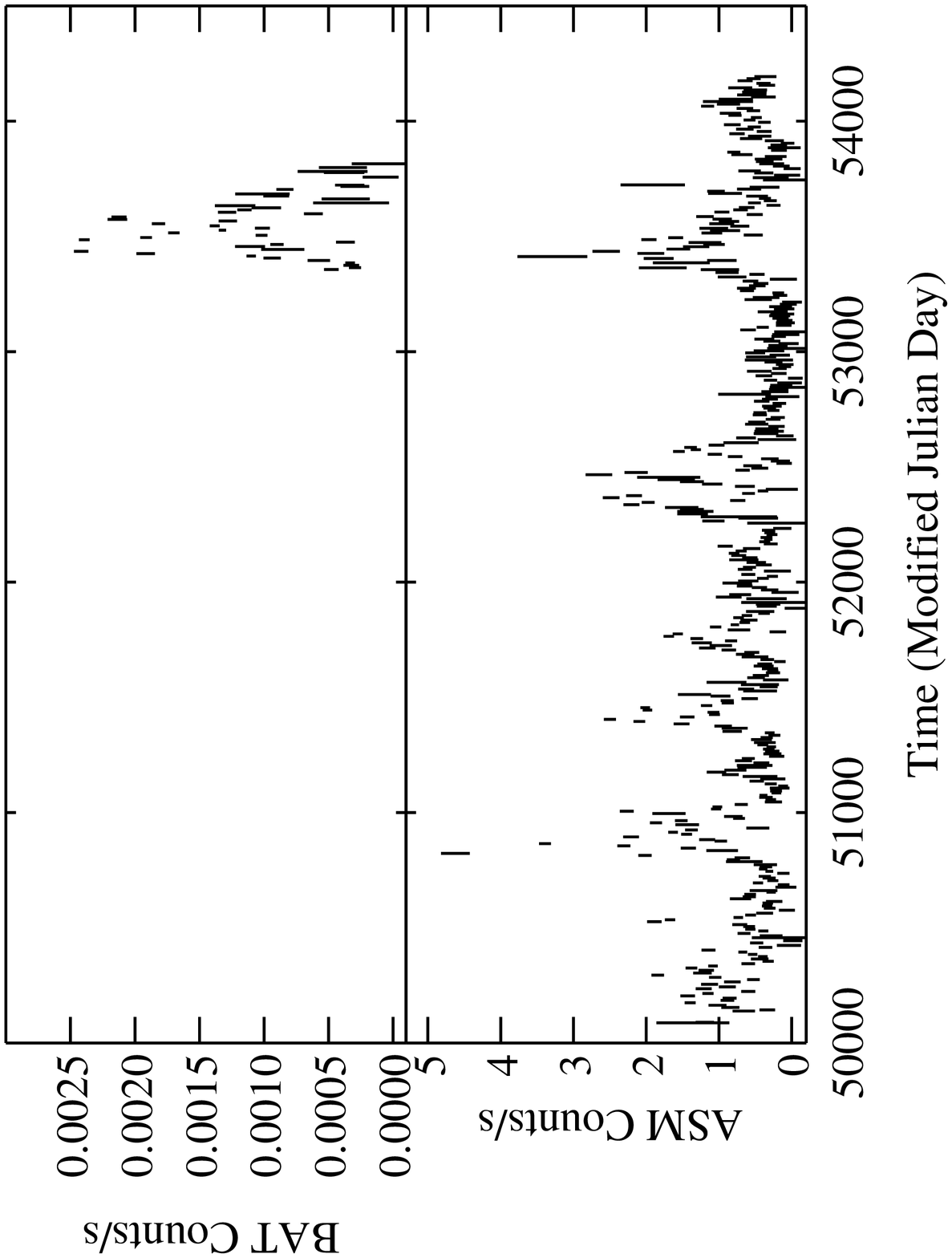}
\figcaption[f7.eps]{BAT and ASM light curves of \src\ in ten day
averages.  The full energy ranges are used for each instrument.
\label{fig:1954_lc1}
}
\end{figure}

\begin{figure}
\epsscale{0.9}
\plotone{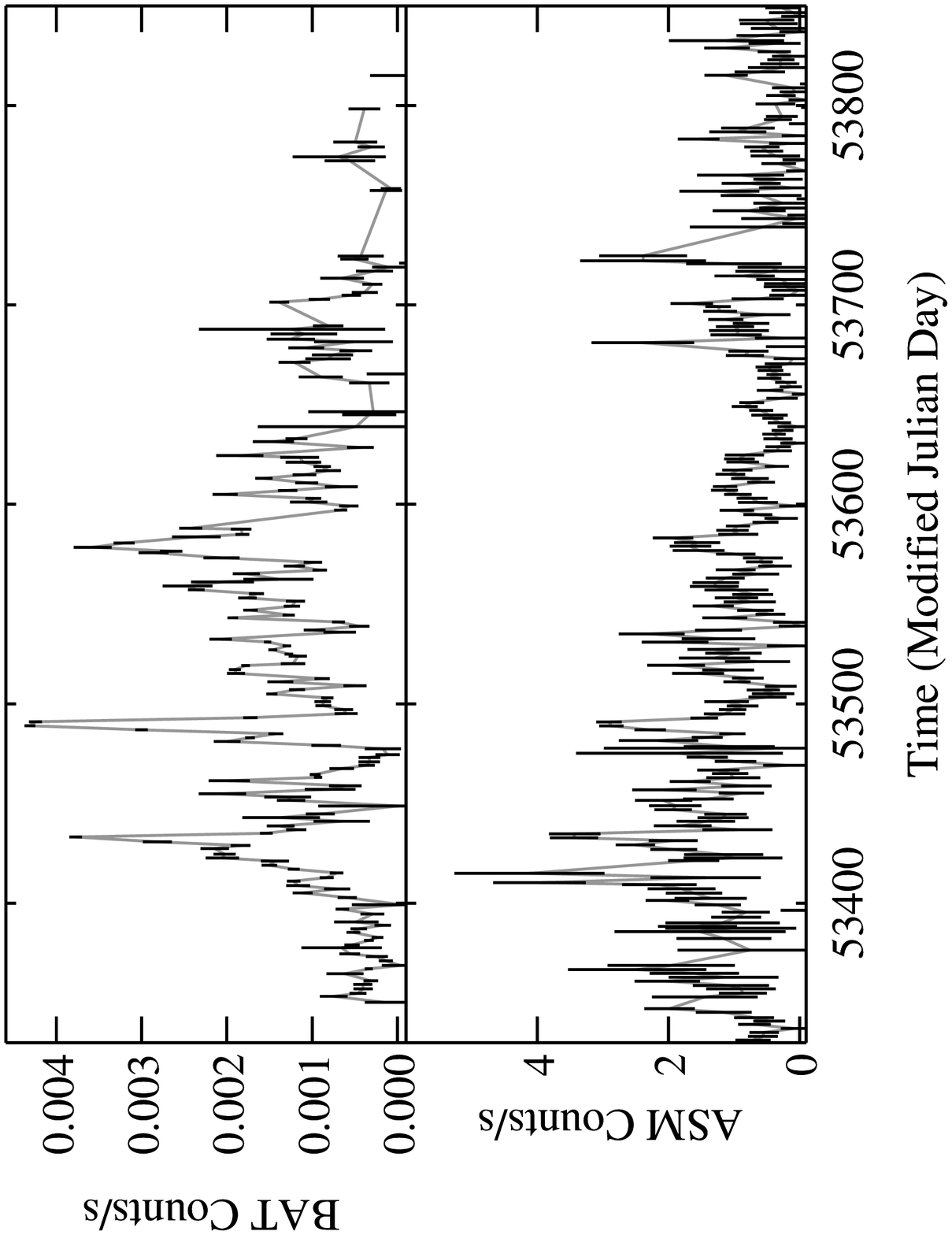}
\figcaption[f8.eps]{Detail of BAT and ASM light curves of
\src\ in three day averages.
The full energy ranges are used for each instrument.
\label{fig:1954_lc2}
}
\end{figure}

The light curves of \src\ obtained with the RXTE ASM and Swift
BAT are shown in Figs. \ref{fig:1954_lc1} and
\ref{fig:1954_lc2} and the mean count rates are given in
Table 1.
The ASM light curve shows the presence of several
outbursts, the most recent of which was also
observed with the BAT. However, some differences
are seen in the structure within the flare (Fig. \ref{fig:1954_lc2}).

\begin{figure}
\epsscale{0.9}
\plotone{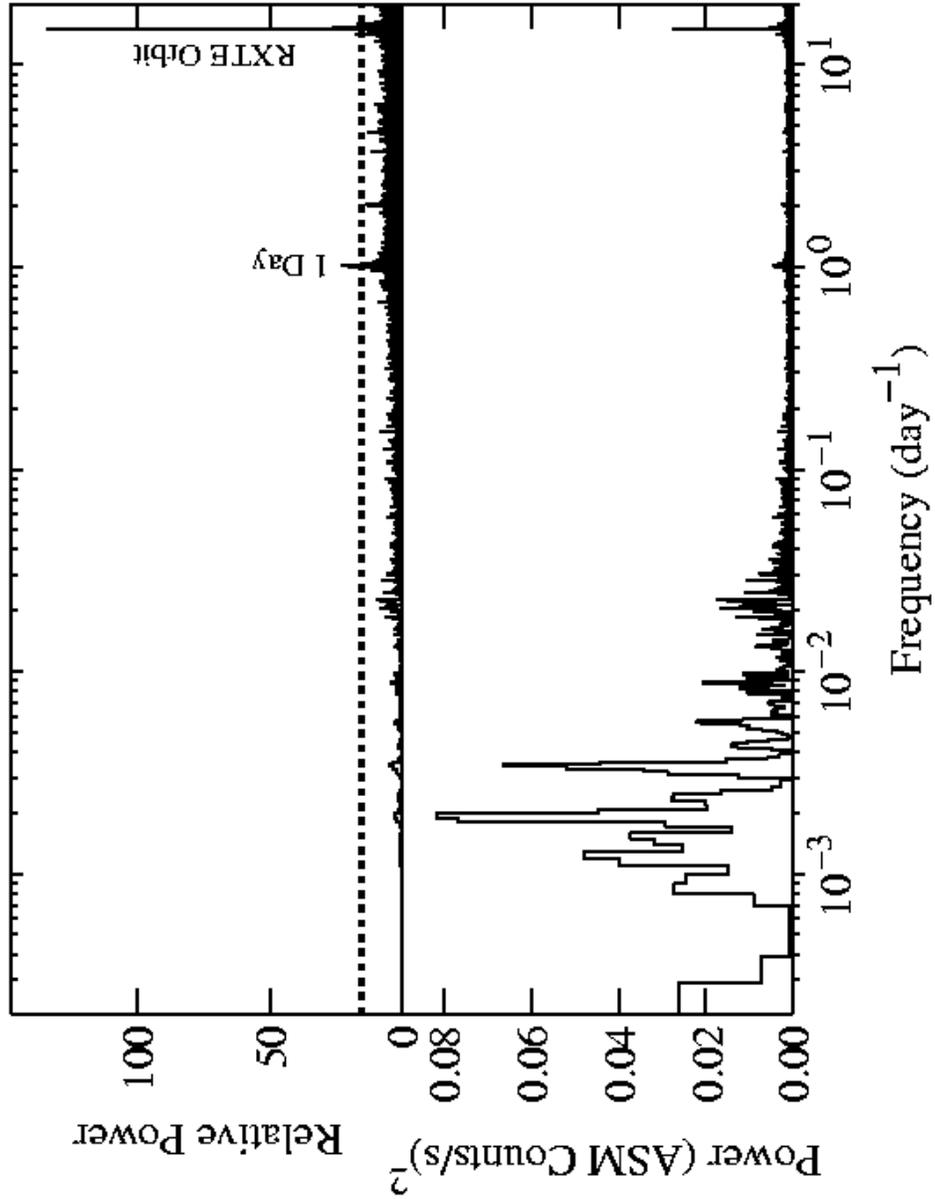}
\figcaption[f9.eps]{Power spectrum of the ASM light curve
of \src.
The lower panel shows the uncorrected power spectrum
and the upper panel shows the power spectrum with the
estimated continuum noise component removed.
The horizontal dashed lines in the upper panel indicate
the estimated 95\% and 99\% significance levels.
\label{fig:1954_asm_ft}
}
\end{figure}

We first searched for long period variations by calculating
the power spectrum of the RXTE ASM light curve. 
We employed the same fitting technique to remove the low-frequency
noise that was employed for \gx.
The resulting power spectra before and after correction
for low-frequency noise are shown in 
Fig. \ref{fig:1954_asm_ft}.
At low frequencies there is no significant detection of
a period that could be due to orbital modulation,
but substantial noise is present in the uncorrected
power spectrum that is likely to be related to
the outbursts that can be seen in the light curve.

We next utilized the more sensitive BAT observations to
investigate higher frequency variability.
This was calculated in the same way as for \gx\ with
the same technique used to remove low-frequency noise.
Due to gaps in the light curve the frequency resolution
was found to be reduced to \sqig3$\times$10$^{-3}$ day$^{-1}$.
The power
spectrum of the entire BAT light curve is shown
in Fig. \ref{fig:bat_ft} with linear frequency scaling to emphasize the
high frequency portion of the spectrum.
The dominant feature in the power spectrum
is a set of peaks around the orbital period of the Swift satellite 
between approximately 14.7 to 15.1 day$^{-1}$.
The peaks seen near 13.92 day$^{-1}$ in \gx, which were
ascribed to a one day$^{-1}$ alias of the Swift orbital period,
are at most only 
very weakly present. This is probably due to the lower flux of
\src\ compared to \gx\ and the correspondingly lower modulation
at the Swift orbital period.
However,
in addition to the features at the Swift orbital
period, a number of other significant peaks
can be seen. Investigation of these peaks shows
that they are all related to a modulation 
at a period of approximately 5.09 hours (4.72 day$^{-1}$)
and aliases with the orbital period of the Swift satellite.
Present are the first and second harmonics of this
\sqig 5 hour period and beats between this period and Swift's orbital
period. The Fourier modulation amplitude at 5 hours is approximately
6.5 $\times$ 10$^{-4}$ \cts.

\begin{figure}
\epsscale{0.9}
\plotone{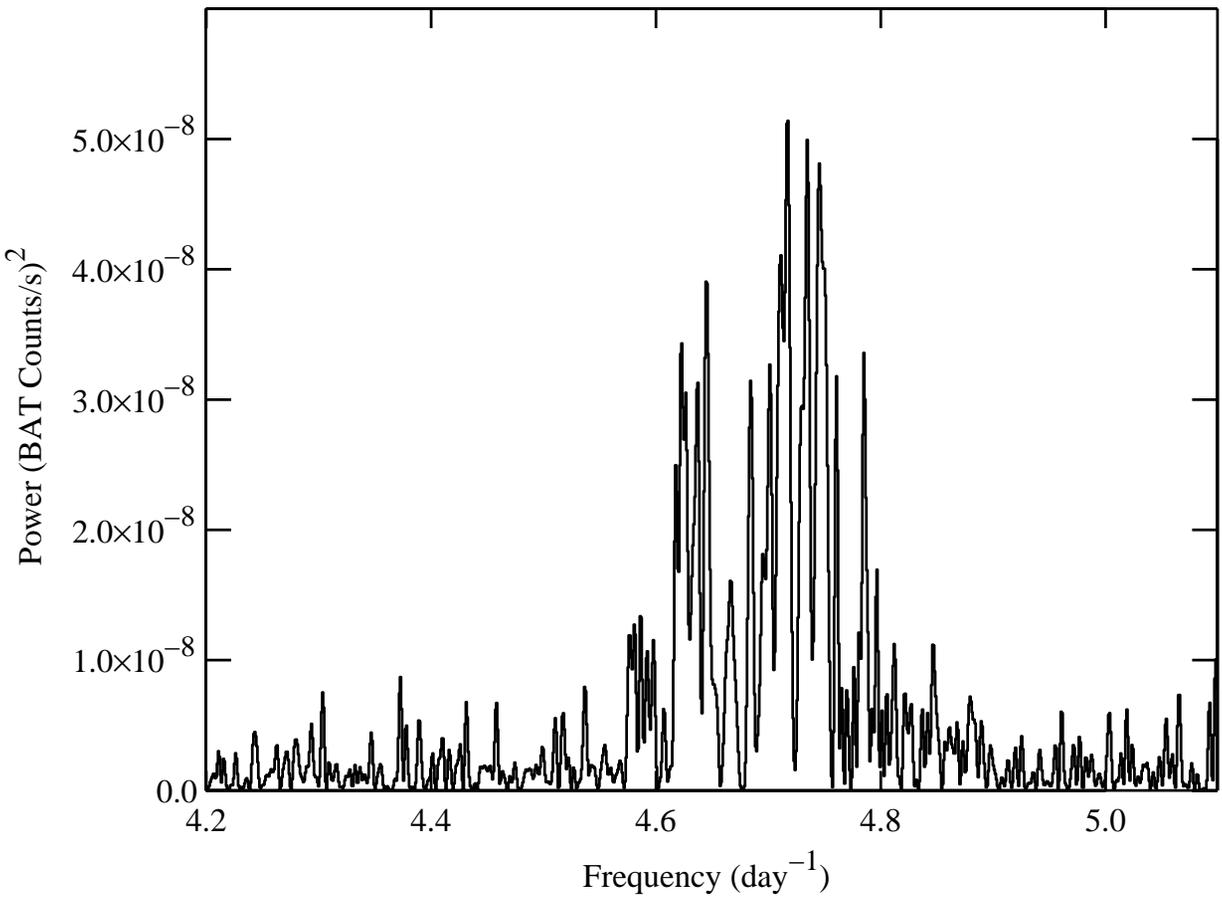}
\figcaption[f10.eps]{Detail of the power spectrum of the BAT light
curve of \src\ around
the 5 hour period. 
\label{fig:bat_ft_detail}
}
\end{figure}

\begin{figure}
\epsscale{0.9}
\plotone{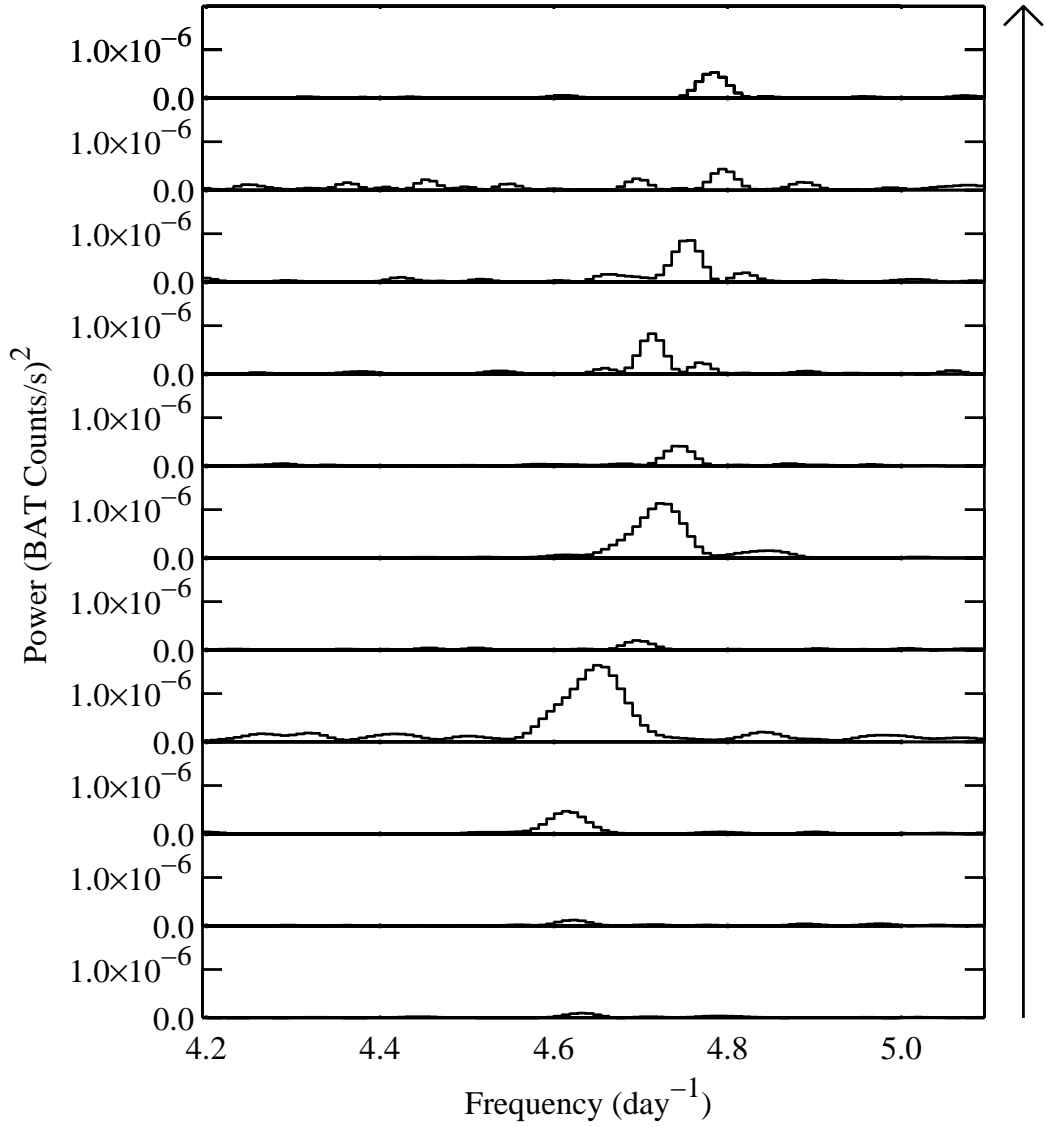}
\figcaption[f11.eps]{Power spectra of BAT observations
of \src. Each power spectrum is obtained from observations
covering 25 days, and each data set is offset by 25
days from the next. Time increases from the bottom panel
to the top.
\label{fig:bat_multi_ft}
}
\end{figure}

\begin{figure}
\epsscale{0.9}
\plotone{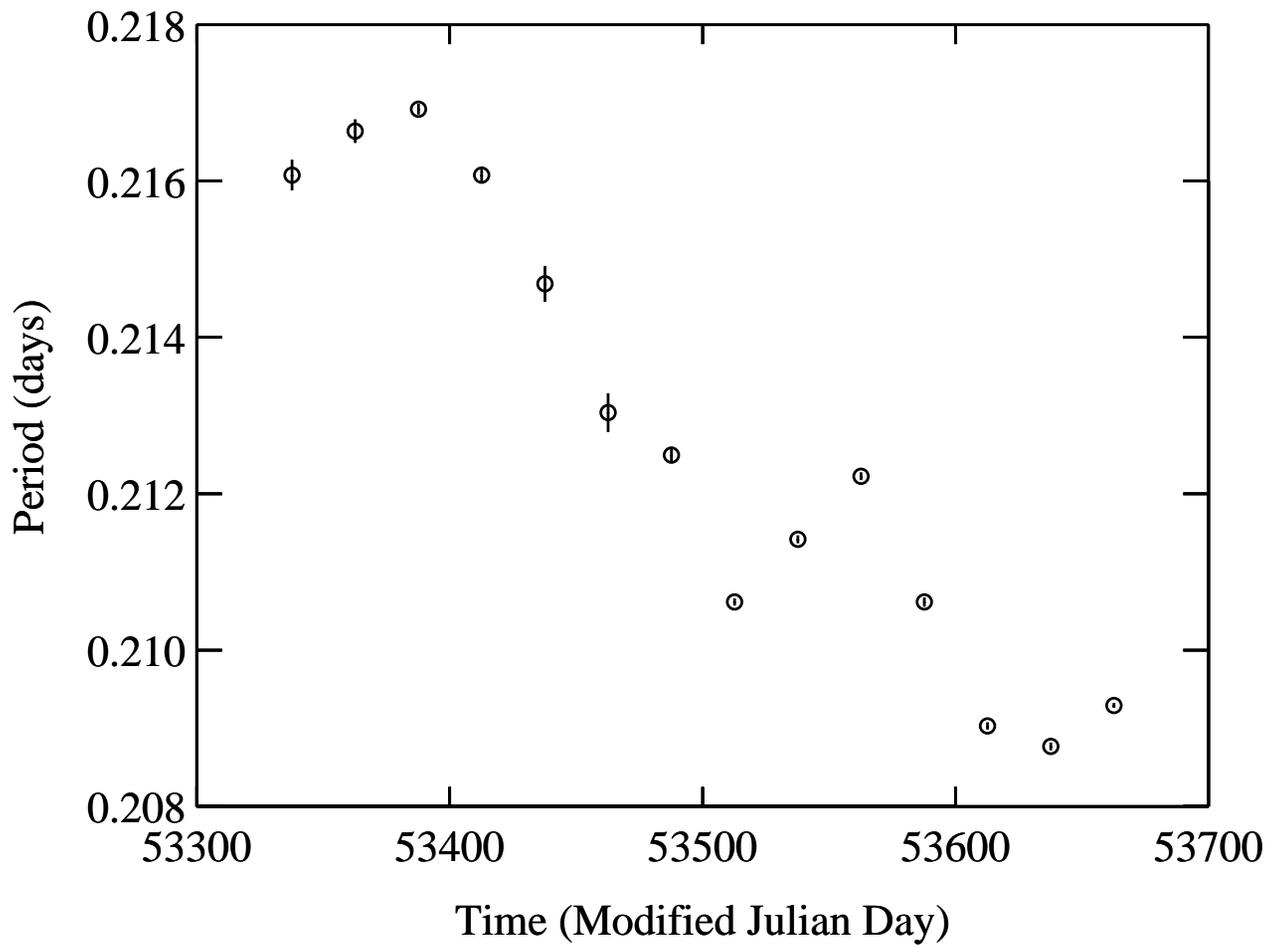}
\figcaption[f12.eps]
{Measurements of the ``pulse'' period of \src.
\label{fig:1954_period}
}
\end{figure}

\begin{figure}
\epsscale{0.9}
\plotone{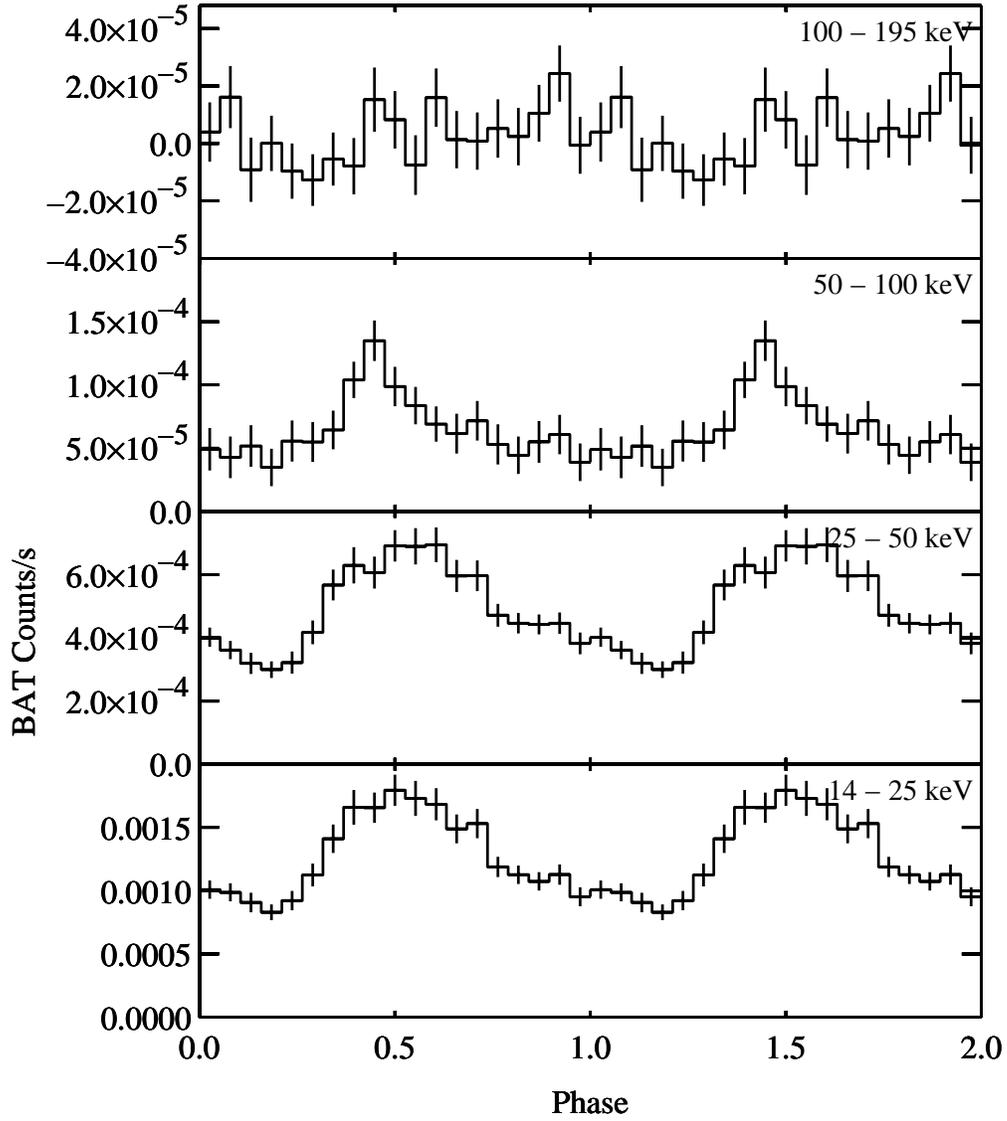}
\figcaption[f13.eps]{BAT light curves of \src\ folded
on the 5 hour period corrected for period changes.
\label{fig:1954_fold}
}
\end{figure}

The peak in the power spectrum
near 5 hours shows a complex structure (Fig. \ref{fig:bat_ft_detail}). 
When power spectra are calculated
from subsets of the BAT light curve (Fig. \ref{fig:bat_multi_ft})
it is seen that during the course of
the BAT observations the period decreased from approximately 5.19 to
5.02 hours in a nearly monotonic fashion (Fig. \ref{fig:1954_period}).
We estimate the period decrease rate to be -2.6 $\pm$ 0.2 $\times$10$^{-5}$
day day$^{-1}$, equivalent to a timescale
of 22 $\pm$ 2 years (see also Mattana et al. 2006).
No trend was seen in the pulsed fraction which had a mean
of \sqig60\%, where the pulsed flux is defined as
the Fourier amplitude divided by the mean flux.
We corrected the BAT light curves for the period change
and folded them on the pulse period 
(Fig. \ref{fig:1954_fold}). The 14 to 25 keV and 25
to 50 keV folded data show similar morphology, whereas the 50 to 100 keV
profile appears to show a maximum at a slightly earlier
phase. The count rate is too low in the 100 to 195 keV band
to measure the pulse profile.

\begin{figure}
\epsscale{0.9}
\plotone{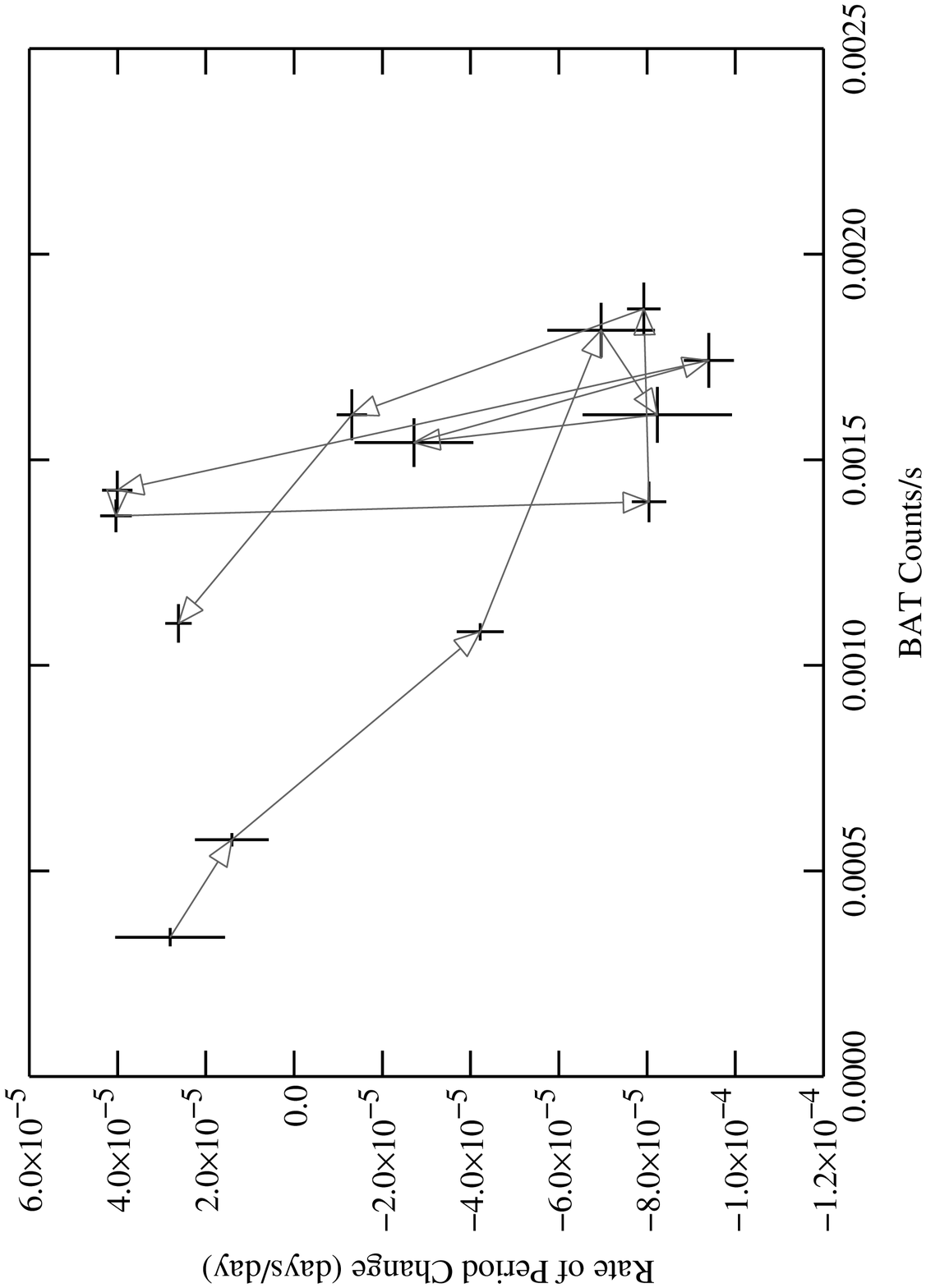}
\figcaption[f14.eps]
{The period change rate for \src\ as a function of
BAT count rate. The arrows indicate the time direction.
\label{fig:1954_pdot_flux}
}
\end{figure}

We next computed the rate of change in the period as
a function of the count rate and this is shown in
Fig. \ref{fig:1954_pdot_flux}. There is a modest anti-correlation
between period change and count rate - the linear correlation
coefficient is -0.53 and the probability of obtaining this
level of correlation from a random data set is 6.5\%.
At the lower count rates at the start and end of the outburst
the rate of period decrease is much lower and might
transition to period increase.

\begin{figure}
\epsscale{0.9}
\plotone{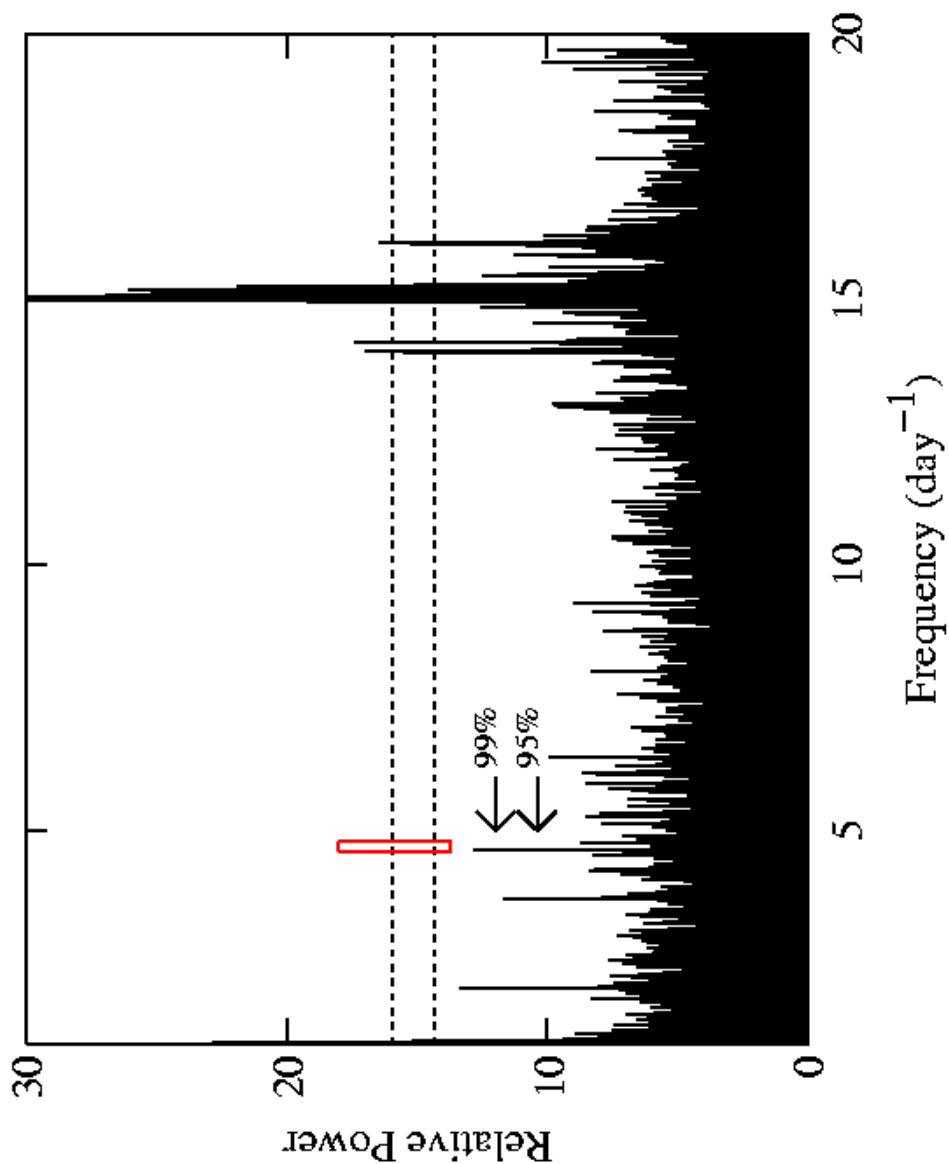}
\figcaption[f15.eps]{Detail of the high frequency portion
of the power spectrum of the ASM light curve of \src.
The range of periods near 5 hours found in the BAT observations
is marked by the red box.
The horizontal dashed lines in the upper panel indicate
the estimated 95\% and 99\% significance levels for the
number of trails in the {\em full} power spectrum.
The arrows show the 95\% and 99\% significance levels
for a period search restricted to between 4.5 to 4.9 day$^{-1}$.
\label{fig:asm_ft_detail}
}
\end{figure}

\begin{figure}
\epsscale{0.9}
\plotone{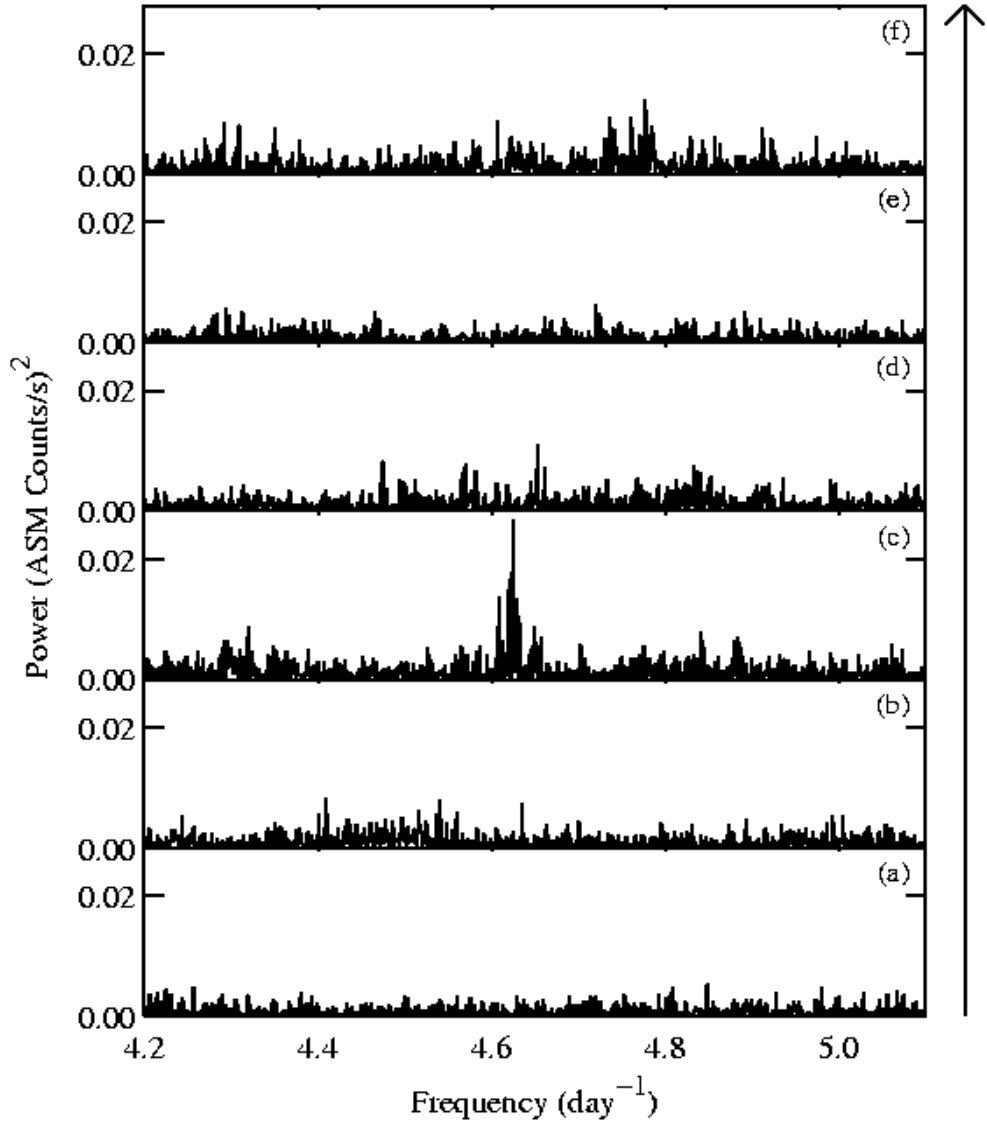}
\figcaption[f16.eps]{Power spectra of the ASM light curve
of \src\ around the 5 hour period. Each power spectrum
is taken from a 655 day long stretch of data
and the time ranges covered are
(a)
MJD  50,087 -
  50,742;
(b)
  50,742 -
  51,397;
(c)
  51,397 -
  52,052;
(d)
  52,052 -
  52,707;
(e)
  52,707 -
  53,362;
and
(f)
  53,362 -
  54,017.
\label{fig:asm_ft_multi}
}
\end{figure}

Because of the detection of the 5 hour period in
the BAT light curve we examined the ASM power spectrum
in more detail
for the possible presence of this period. The ASM power spectrum
is plotted with linear frequency scaling in Fig. \ref{fig:asm_ft_detail}.
A low amplitude peak is seen near 4.72 day$^{-1}$, consistent
with the range of frequencies found in the BAT.
If we make a significance calculation for the ASM peak
with an assumed restricted range of frequencies of 4.5 to 4.9 day$^{-1}$,
which encompasses the range of frequencies seen with the BAT,
then the ASM detection is significant at $>$ 99\%.
We then investigated dividing the ASM power light curve into sections
and calculating power spectra of these
to investigate the time dependence of this peak.
After some experimentation, we found that 
dividing the light curve into 655 day duration sections
gave a reasonable balance between having sufficient data in
a section to show a peak, and providing information on
when this modulation near the 5 hour period may have occurred.
Fig. \ref{fig:asm_ft_multi} shows the ASM power spectrum in the
4.2 to 5.1 day$^{-1}$ range for the light curve divided
into these sections, the start and end times of
the light curves are given in the Figure caption. 
For this choice of time ranges
there is only one possible detection of modulation near the 5
hour period. This is for the section
of the light curve that was obtained between MJD 51,397 and
52,052.

\begin{figure}
\epsscale{0.9}
\plotone{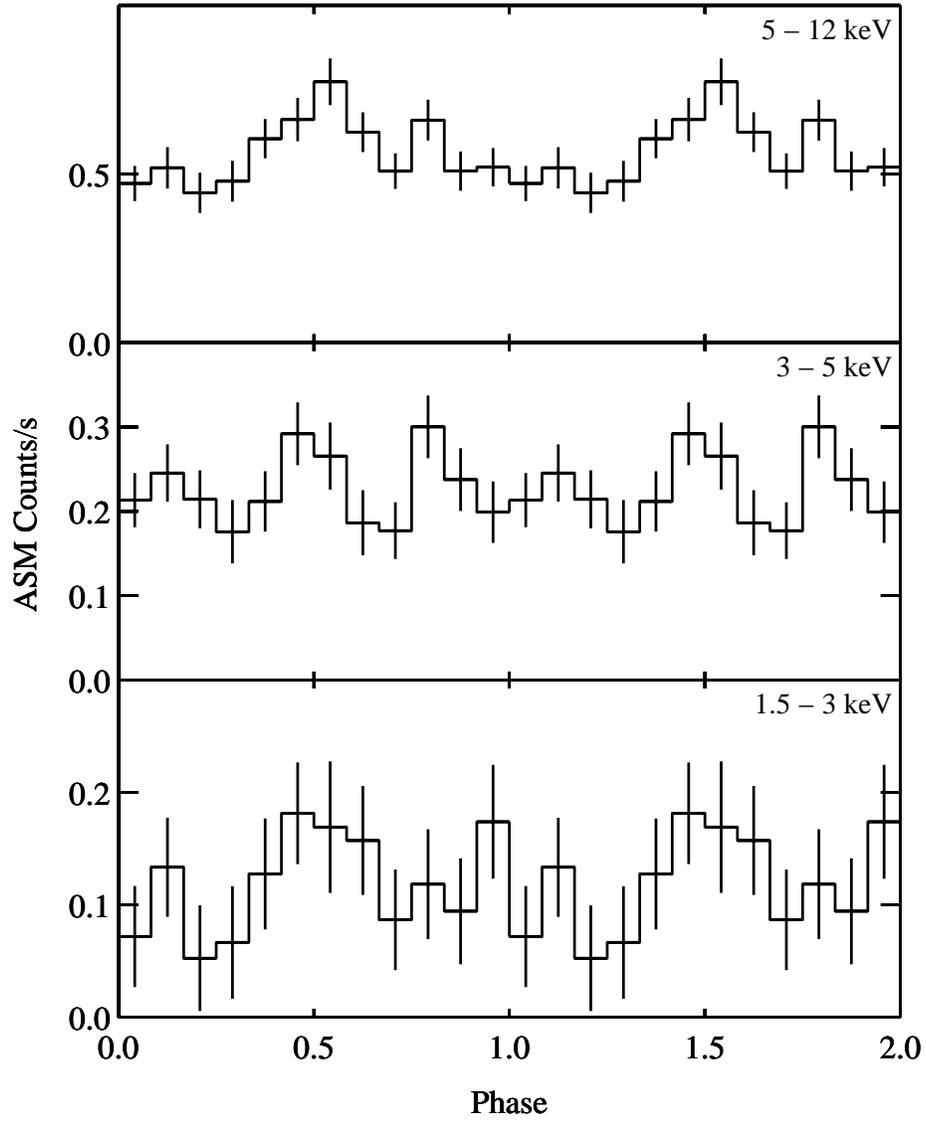}
\figcaption[f17.eps]{ASM light curves of \src\ from
the time range covered by BAT observations folded
on the 5 hour period corrected for period changes in the same
way as for Fig. \ref{fig:1954_fold}.
\label{fig:asm_1954_fold}
}
\end{figure}

We next corrected the ASM light curves in the time range MJD
53,350 to 53,650 for the period changes detected in
the BAT light curve, and folded the energy separated data
(Fig.  \ref{fig:asm_1954_fold}). The folded 1.5 - 3 keV
and 3 - 5 keV bands
do not exhibit any obvious modulation. The folded 5 - 12 keV 
ASM light curve suggests a possible modulation which peaks
at the same phase as the folded BAT light curves.

\subsection{\17}

\begin{figure}
\epsscale{0.9}
\plotone{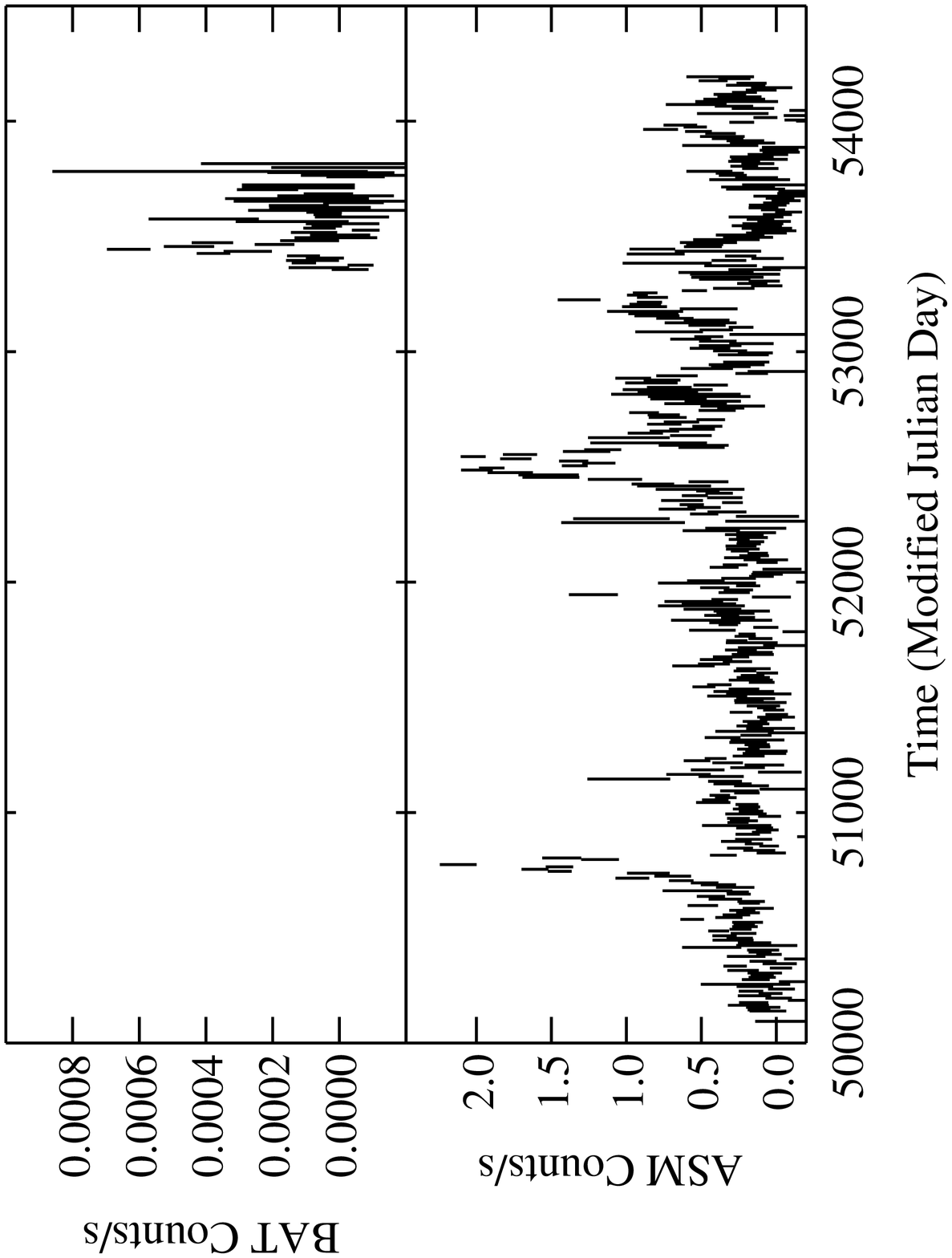}
\figcaption[f18.eps]{BAT and ASM light curves of \17\ in ten day
averages.  The full energy ranges are used for each instrument.
\label{fig:1700_lc}
}
\end{figure}

The ASM and BAT light curves of \17\ are shown
in Fig. \ref{fig:1700_lc} and the mean count rates
are given in Table 1. 
The ASM light curve shows considerable variability
with at least two prominent outbursts.

\begin{figure}
\epsscale{0.9}
\plotone{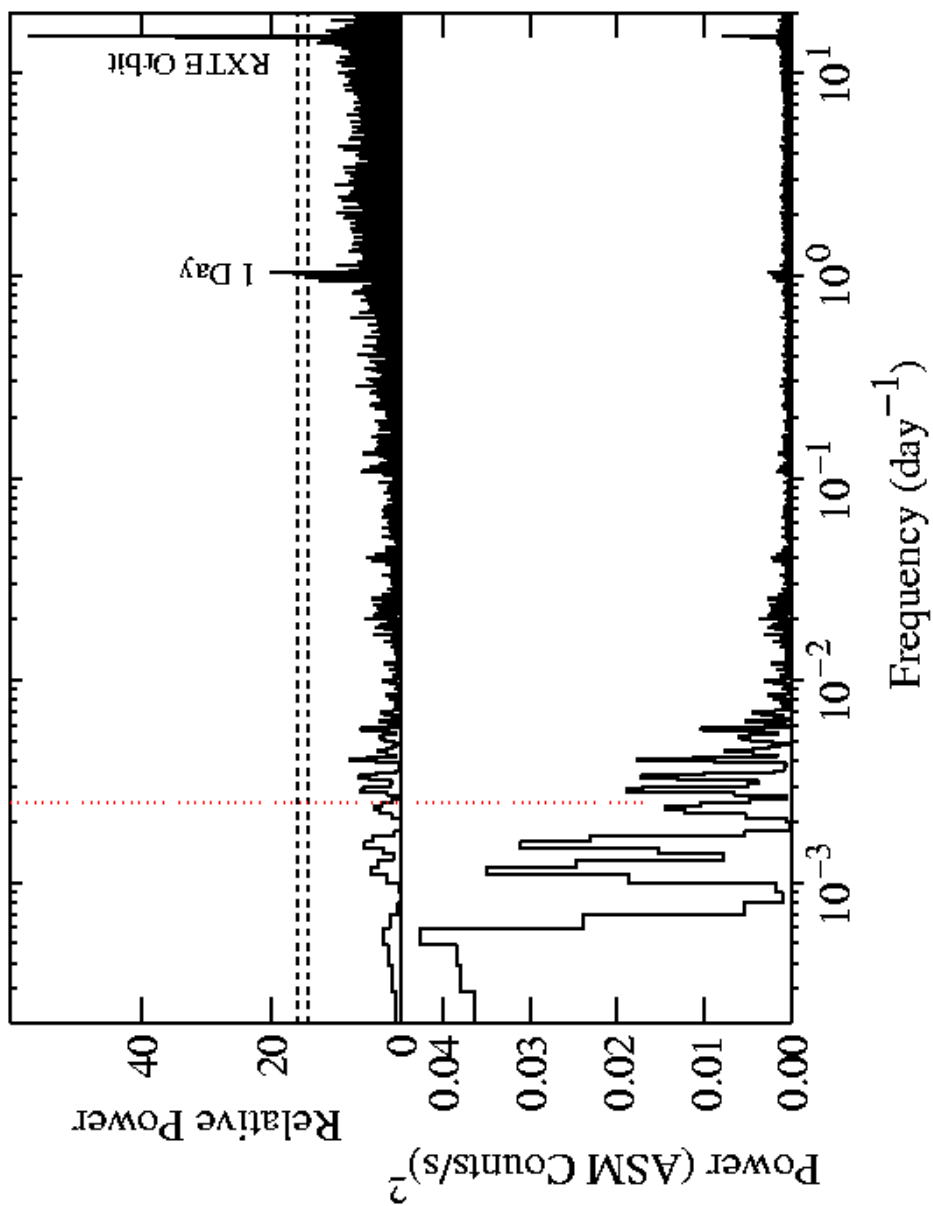}
\figcaption[f19.eps]{Power spectra of the ASM light
curve of \17. 
The lower panel shows the uncorrected power spectrum
and the upper panel shows the power spectrum with the
estimated continuum noise component removed.
The horizontal dashed lines in the upper panel indicate
the estimated 95\% and 99\% significance levels.
The vertical dotted red line marks
the 404 day period previously reported for this source
by Masetti et al. (2002) and Galloway et al. (2002).
\label{fig:1700_asm_ft}
}
\end{figure}

\begin{figure}
\epsscale{0.8}
\plotone{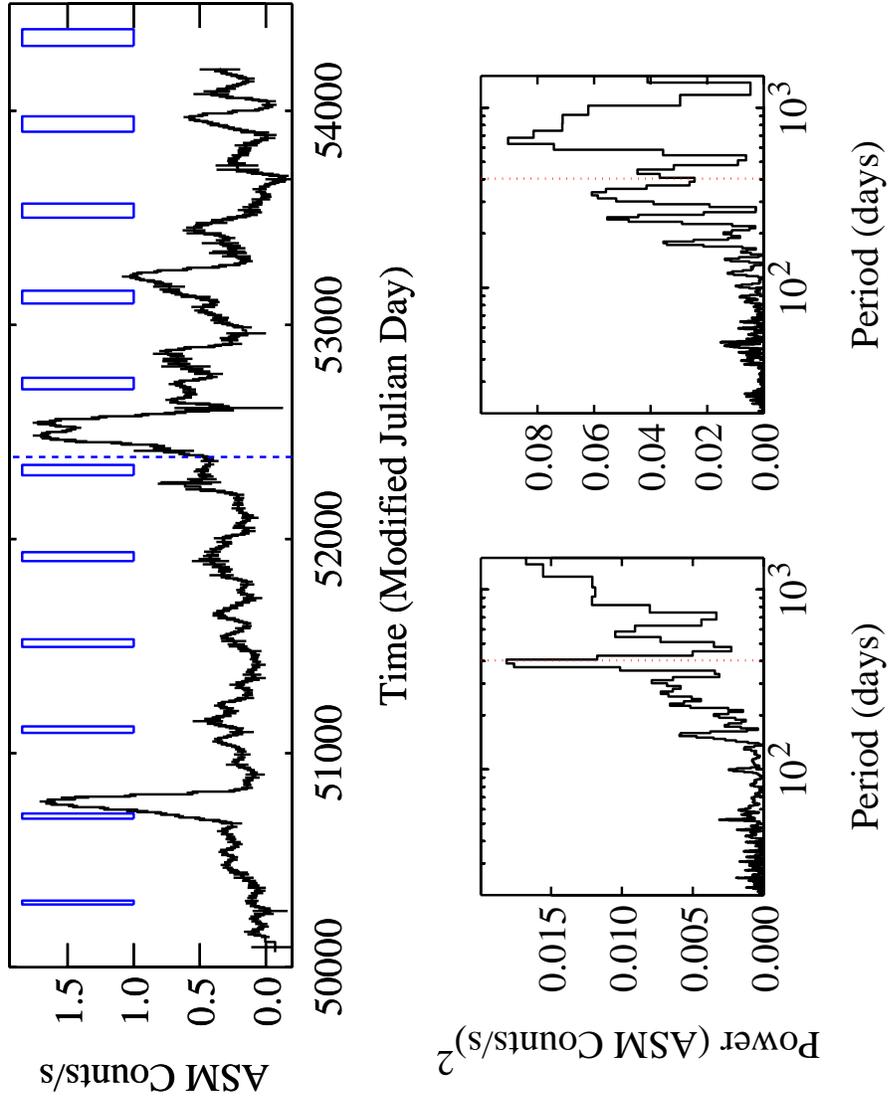}
\figcaption[f20.eps]{Top panel: The ASM light curve of \17.
The times of predicted periastron passage using the
ephemeris of Galloway et al. (2002) are shown
by the blue boxes. The width of the boxes indicates
the error due to the 3 day error on the orbital period
obtained by Galloway et al. The 80 day error on the
epoch of periastron passage is not included.
Data earlier than the time marked by the vertical dashed line
(MJD 52383) were included in Galloway et al. (2002).
Bottom left panel: the power spectrum of the 
light curve of \17\ between MJD 50,087 to 52,383.
Note that the x-axis is in units of period 
to facilitate comparison with Fig. 2 of Galloway et al.
(2002). The vertical dotted red line marks
the 404 day period reported by Galloway et al.
Bottom right panel: the power spectrum of the light
curve between MJD 52,383 to 54,195.
\label{fig:1700_asm_ephem}
}
\end{figure}

The power spectrum of the ASM light curve of \17\ was
obtained in the same way as for \gx\ and \src\ including
the same removal of low-frequency noise.
The uncorrected and corrected power spectra are shown
in Fig. \ref{fig:1700_asm_ft} where the dashed line
indicates the 404 day period
proposed by
Masetti et al. (2002) and Galloway et al. (2002).
This period does not coincide with any significant peak
in the power spectrum. Although a small peak is
present near this period in the uncorrected power spectrum
this is apparently part of the low-frequency noise.
In Fig. \ref{fig:1700_asm_ephem} (top panel) we plot a smoothed
version of the ASM light curve together with the predicted
times of periastron passage using the orbital fit
of Galloway et al. (2002). The division between
the data presented in Galloway et al. (2002) and data
obtained subsequently is marked by the dashed line.
Although some flares coincide with the predicted
times of periastron passage, many do not.
In particular, the largest flare in the light curve
around approximately MJD 52500 is near the time
of predicted apastron passage.
In the bottom panels of Fig. \ref{fig:1700_asm_ephem} we plot
the power spectra of the two light curve sections, {\em not} corrected
for low frequency noise. The power spectrum of the
later section of the light curve does not show modulation at
404 days.

The BAT light curve of \17\ shows that the source was in a low
luminosity state during this time.
We corrected the power spectrum for any low-frequency
components as for \gx\ and \src. However, Fig. \ref{fig:bat_noise_fits}
shows that the power spectrum was fairly flat compared to
the other two sources.
The corrected power spectrum 
of the BAT light curve is plotted in Fig. \ref{fig:bat_ft}. 
It does not show
the presence of any periodicities from the source at high
frequencies and the only significant feature present is at the Swift
orbital period. The one day$^{-1}$ alias peaks of the Swift
orbital period
near 13.92 day$^{-1}$ seen in \gx\
are not seen in \17, which is likely due to the relative
faintness of \17\ and the low amplitude of the modulation
at the Swift orbital period.

\section{Discussion}

\subsection{Lack of Orbital Flux Modulation}

None of the three symbiotic X-ray binaries considered here shows
modulation on long timescales which could be interpreted
as the orbital period of the system. There is no modulation
at the 1161 day orbital period of GX 1+4 and the previously
proposed modulation period of 404 days in \17\ is also
not a persistent feature of the X-ray
light curve. The long term light curves of all three
of these systems appear to be dominated by irregular
flaring type behavior. 
Fig. \ref{fig:noise_fits} illustrates the importance of
removing low-frequency noise before searching for the long
orbital periods expected in this type of system.
The lack of such subtraction probably accounts for
the previously reported periods in these sources.
Although Galloway et al. (2002) also reported radial
velocity variations that appeared to coincide with
the claimed X-ray period, the significance of the
period was low, and these authors also reported evidence
for radial velocity variations on longer timescales.
The cause of the flares in these sources
may be variability in the stellar wind of the M
giant as almost all red giants are 
optically photometrically variable on time scales of
hundreds to thousands of days (e.g. Percy et al. 2001).
If the mass loss rate in the wind
varies in conjunction with the photometric variations
then this could
produce a modulation of the accretion rate onto the neutron star.
For GX 1+4, a modest eccentricity of 0.10$\pm$0.02 was reported by
Hinkle et al. (2006).
Our results show that for \gx, \src, and \17\
any accretion rate modulation caused
by an eccentric orbit is 
much less than that caused by other mechanisms.

\subsection{The 5 hour period in \src}

Masetti et al. (2006)
determined the optical counterpart of 4U 1954+31 to be an M 4-5
giant which would imply a long orbital period in order for the
star not to overfill its Roche lobe. For
comparison, Chakrabarty \&
Roche (1997) derive a constraint of P$_{orb}$ $>$ 100 days
for GX 1+4 where the donor is an M5 III star. 
The 5 hour period thus cannot be the orbital period of
a compact object about the M star.
The change in the 5 hour period cannot be predominantly due
to orbital Doppler modulation. If this was the case
from Fig. \ref{fig:1954_period} 
it can be seen that the orbital period would be $\gtrsim$ 500 days, and the
velocity semi-amplitude would be $\gtrsim$ 10,000 \kms.
These values would give a huge 
mass function of $\gtrsim$5$\times$ 10$^6$M\sun.
The large change in the 5 hour period must thus
be intrinsic to the source of the modulation. This makes
it implausible that this 5 hour period is
caused by orbital modulation which rules out triple star
models.
We note that the period change is also too
large for the period to come from the rotation period of an accreting white
dwarf because of the large moment of inertia of these objects.
We therefore consider other possible models
that could explain the 5 hour modulation.

\subsubsection{Neutron Star Rotation Period}

The 5 hour period in \src\ is somewhat reminiscent of the 2.7 hour period seen
in the high-mass X-ray binary 2S 0114+650.
This period was discovered by Finley, Belloni, \&
Cassinelli (1992) and was proposed to be a
neutron star rotation period. Apparent confirmation
of this interpretation comes from the continued
presence of the 2.7 hour modulation 
in the ASM light curve (Corbet et al.
1999; Farrell et al.  2006) which implies 
the modulation is rather coherent. An indirect argument
for the modulations on periods of hours in both \src\
and 2S 0114+650 being neutron star pulsation periods
is that no other periodicity has been found in
these systems that could be neutron star pulsation
periods.
A long period of 1.6 hours has also been reported
for IGR J16358-4726 (Kouveliotou et al. 2003,
Patel et al. 2004). However, for IGR J16358-4726 it is not clear
whether this period is a neutron star rotation period
or an orbital period.
A 5 hour neutron star rotation period for \src\ would make it one
of the longest known with only the 6.67 period from the point source
in the supernova remnant RCW 103 exceeding this value
(De Luca et al. 2006).

If the periodicity in \src\ does 
indeed represent the rotation period of a neutron star,
then the rapid spin-up requires an X-ray luminosity of approximately
5$\times$10$^{35}$ \ergss\ (e.g. Mason 1977,
Joss \& Rappaport, 1984). 
For comparison, in an earlier outburst of \src\ Masetti et al. (2007a)
derive luminosities of \sqig2$\times$10$^{35}$ \ergss, {\em not} corrected
for absorption from RXTE and BeppoSAX observations
obtained on MJD 50796 and 50937 respectively. Masetti et al. (2007a)
also found the absorption to be complex and variable.
The observed X-ray flux therefore appears to be consistent
with that required to account for the observed spin-up.

For a 5 hour rotation period, material corotating in a Keplerian orbit
about the neutron star would be at a radius of \sqig10$^{11}$ cm.
In contrast, the magnetospheric radius, at which the magnetic
field would dominate over the ram pressure, would be
located at 
\sqig 10$^{9}$ (L$_X$/10$^{35}$\ergss)$^{-2/7}$ (B/10$^{12}$G)$^{4/7}$ cm.
The 5 hour period is thus apparently much longer than the
equilibrium spin period. This appears to be similar to
the case of GX 1+4 where the corotation radius is estimated
to be 3 $\times$ 10$^{9}$ cm and the magnetospheric radius is estimated to
be significantly smaller at 
\sqig 3 $\times$ 10$^{8}$ cm for a magnetic field of
10$^{12}$ G (Chakrabarty \& Roche 1997).
Despite this, GX 1+4 has primarily exhibited long-term spin-down,
with occasional spin-up behavior seen in the brightest
states (e.g. Chakrabarty et al. 1997, Ferrigno et al. 2007).
In order to explain the spin-down in \gx\ without the need
for an unusually strong magnetic field,
it has been proposed (e.g. Makishima et al.
1988, Chakrabarty et al. 1997, Nelson et al. 1997) that a counter-rotating
accretion disk may sometimes form.
Alternatively, Perna et al. (2006) consider a spin-down mechanism
related to the fallback of material expelled when a neutron
star is in the propeller regime.

The possible detection of the 5 hour period in \src\ with the ASM
was at a similar period to that seen at the start of the outburst observed
with the BAT. This suggests that \src, like \gx, may also exhibit
spin-down between outbursts that maintains the period
at this length.
The possibility of spin-down is also suggested by the lower flux
points in Fig. \ref{fig:1954_pdot_flux}.
The dependence of $\dot{P}$ on flux is predicted by standard accretion
models (e.g. Ghosh \& Lamb 1979) and has been seen, for
example, in the HMXB EXO 2030+375 (Parmar et al. 1989).
The period change results appear to suggest that
\src\ may be rotating close to its equilibrium period, and hence
that the neutron star has a very large magnetic field ($\sim$10$^{15}$ G). 
However,
the complex $\dot{P}$ behavior exhibited by
\gx\ indicates that caution should be taken with this
interpretation 
(e.g. Chakrabarty et al. 1997).
Additional observations are
needed to confirm whether \src\ does indeed consistently show spin-down
during periods of low flux and more completely quantify 
the $\dot{P}$ dependence on source luminosity.


\subsubsection{Accretion Instability}

The 5 hour modulation might alternatively be caused by some type of
instability in the accretion flow onto the neutron star.
The 5 hour timescale is similar to that of
the 3.96 hour quasi-periodic flares observed on one occasion from the
Be/neutron star binary EXO 2030+375 (Parmar et al. 1989). As this
modulation was only detected once for a brief period we have no
information on whether such a modulation might exhibit long term
period changes such as shown by \src.  Therefore, although this model
cannot yet be definitely excluded, its transient appearance in only
one other source may make it a less likely explanation.

\subsubsection{Variations in the Mass Donor}
The interpretation of the 2.7 hour period in 2S 0114+650 as
a neutron star rotation period has been questioned
by Koenigsberger et al. (2006) who propose that the
period comes from tidal interactions which drive oscillations
in the B supergiant primary. However, an explanation of the 5 hour
period in \src\ as
a pulsation in the mass donor appears problematic as the fundamental
pulsation periods of M giants are very long and there is little
evidence for pulsations on periods less than 10 days
(Percy et al. 2001, Koen et al. 2002).
We therefore exclude this model as a possible explanation of
the 5 hour periodicity.

\subsection{Summary of Models}
The persistence of the 5 hour period over an extended length of time,
the change in period consistent with accretion torque, and lack of any
other periodicity in the light curve that could be a pulse period, all
suggest that the 5 hour period is a neutron star rotation
period. However, an EXO 2030+375 type instability is not yet
completely ruled out.

\section{Conclusion}

The three symbiotic X-ray binaries considered here
show substantial X-ray
variability on long time scales, dominated by the
presence of slow flares.
In none of the systems
does
the long term X-ray flux show detectable orbital modulation
and there is no evidence for modulation at any of
the previously reported periods. Any accretion
rate variations caused by modulation in an eccentric
orbit appear to be much less than that which occurs
during the non-periodic flares.

In contrast, on shorter timescales two of the three sources, \gx\ and
\src, exhibit periodic modulation that appears to come from
a neutron star rotation period. This, together with
the pulsations shown by Sct X-1 and 1RXS J180431.1-273932, suggests that 
further observations of \17, preferably during an outburst,
might also reveal a pulsation period in this source.
Additional observations of \src\ should help to
better constrain the nature of the 5 hour modulation
by showing whether this is a persistent feature of
the light curve and, if so, how the period changes
on long timescales in response to varying accretion rates.

\acknowledgements
We thank the referee for useful comments.
JLS acknowledges support through grant SAO G06-7022AA.

\begin{deluxetable}{lcccccccc}
\rotate
\tablewidth{8.2in}
\setlength{\tabcolsep}{0.02in} 
\tabletypesize{\scriptsize}
\tablecaption{Symbiotic X-ray Binaries}
\tablehead{
\colhead{System} & \colhead{P$_{spin}$} & \colhead{P$_{orb}$}
& \colhead{Optical} & \colhead{Spectral} & \colhead{Distance}
& \colhead{Mean ASM} & \colhead{Mean BAT} \\
\colhead{} & \colhead{} & \colhead{(days)}
& \colhead{Counterpart} & \colhead{Type} & \colhead{(kpc)}
& \colhead{Count Rate} & \colhead{Count Rate}
}
\startdata
GX 1+4 & 120 s (C97) &  1161 (H06), [304 (P99)] & V2116 Oph (GF73) & M5 III 
     & 3-6 (CR97) &  1.1  &  1.4$\times$10$^{-3}$ \\
\src\ &  5 - 5.2 hr (C07) & ? &  U1200\_13816030 (M06) & M4 III
    & 1.7 (M06) &  0.7 & 1.1$\times$10$^{-3}$ \\
\17 & [900s (MG01)] & [404 (G02)] &  HD 154791/V934 Her (M06) &  M2 III 
   & 0.42 (M02) &  0.4 & 9$\times$10$^{-5}$ \\
\tableline
Sct X-1 &  111 - 113s (K07) & ? & -- & late K/early M I-III
     &  $\geq$ 4 (K07)  &  0.2 & --\\
IGR J16194-2810 & ? & ? &  U0600\_20227091 (M07) &  M2 III
    &  $\leq$ 3.7 (M07) & -- & -- \\
1RXS J180431.1-273932 & 494 (N07) & ? & OGLE II DIA BUL-SC35-4278  &
M6III? & 10? (N07) & -- & --\\ 

\enddata
\tablecomments{
(i) Parameters in square brackets are considered to be
questionable.
(ii) References: 
C07 - this work,
C97 - Chakrabarty et al. (1997),
CR97 - Chakrabarty \& Roche (1997),
G02 - Galloway et al. (2002),
GF73 - Glass \& Feast (1973),
H06 - Hinkle et al. (2006),
K07 - Kaplan et al. (2007),
M02 - Masetti et al. (2002),
M06 - Masetti et al. (2006),
M07 - Masetti et al. (2007b),
MG01 - Morgan \& Garcia (2001),
N07 - Nucita et al. (2007),
P99 - Pereira et al. (1999).
References are not exhaustive and additional
references can generally be found within those given.
The same references apply to both distance and spectral type.
} 
\end{deluxetable}


\clearpage

\clearpage

\clearpage

\clearpage

\clearpage

\clearpage

\end{document}